\documentclass[a4paper,11pt]{article}
\pdfoutput=1
\usepackage[text={17.1cm,24.6cm},centering]{geometry}
\usepackage[utf8]{inputenc}
\usepackage{physics}
\usepackage{amsmath,amssymb,mathtools,color}
\usepackage{subfigure}
\usepackage{pst-node}
\usepackage{tikz-cd} 
\usepackage{amssymb}
\usepackage{mathrsfs} 
\usepackage{graphicx}
\usepackage{authblk,floatrow}
\usepackage{cite}
\usepackage[colorlinks=true,linkcolor=blue, citecolor=red, urlcolor=blue, bookmarks]{hyperref}
\numberwithin{equation}{section}

\title{Entanglement of the $3$-State Potts Model via Form Factor Bootstrap: \\
Total and Symmetry Resolved Entropies}
\author[1]{Luca Capizzi\footnote{email: lcapizzi@sissa.it}}
\author[1]{D\'avid X. Horv\'ath}
\author[1,2]{Pasquale Calabrese}
\author[3]{Olalla A. Castro-Alvaredo}
\affil[1]{SISSA and INFN, Via Bonomea 265, 34136 Trieste, Italy}
\affil[2]{International Centre for Theoretical Physics (ICTP), Strada Costiera 11, 34151 Trieste, Italy}
\affil[3]{Department of Mathematics, City, University of London, 10 Northampton Square, EC1V 0HB, London, UK}
\date{}                     
\def \be {\begin{equation}} 
\def \ee {\end{equation}} 

\def \l {\left(} 
\def \r {\right)} 
 
\def \la {\left\langle} 
\def \ra {\right\rangle} 
\def \beq {\begin{equation}}
\def \eeq  {\end{equation}}
\def \beqa {\begin{eqnarray}}
\def \eeqa  {\end{eqnarray}}
\def \TT     {\mathcal{T}}

\begin{document}
\maketitle
\begin{abstract}
In this paper, we apply the form factor bootstrap approach to branch point twist fields in the $q$-state Potts model for $q\leq 3$.  For $q=3$ this is an integrable interacting quantum field theory with an internal discrete $\mathbb{Z}_3$ symmetry and therefore provides an ideal starting point for the investigation of the symmetry resolved entanglement entropies. However, more generally, for $q\leq 3$ the standard R\'enyi and entanglement entropies are also accessible through the bootstrap programme. In our work we present form factor solutions both for the standard branch point twist field with $q\leq 3$ and for the composite (or symmetry resolved) branch point twist field with $q=3$. 
In both  cases, the form factor equations are solved for two particles and the solutions are carefully checked via the $\Delta$-sum rule. Using our analytic predictions, we compute the leading finite-size corrections to the entanglement entropy and entanglement equipartition for a single interval in the ground state.
\end{abstract}

\section{Introduction}

Symmetries play a crucial role in different areas of modern physics, e.g. they are the key concept in the formulation of the standard model of fundamental particles. Their interplay with entanglement properties in the context of extended quantum systems has been however pointed out just in the last years. Before entering into the core of the subject, we briefly  recall that when a system is described by a pure state, the entanglement content of a subsystem $A$ may be quantified by the $n$-th R\'enyi entropy \cite{vNE1,vNE2,vNE3,vNE4}
\be
S_n = \frac{1}{1-n}\log \text{tr}\l \rho_A^n\r,
\ee
where $n$ is a positive real number and $\rho_A$ is the reduced density matrix of the subsystem $A$. From those, the von Neumann entropy is obtained by taking the limit $n\rightarrow 1$
\be
S_1 \equiv \underset{n\rightarrow 1}{\lim} S_n  = -\text{tr}\l\rho_A\log \rho_A\r.
\ee
It is worth saying that the knowledge of $S_n$, for $n$ integer $\geq 2$, gives all the information about the set of eigenvalues of $\rho_A$ \cite{ace-18,cl-08}.\\
The idea of the symmetry resolution of entanglement consists in the understanding of how the internal structure of entanglement is related to a certain symmetry of the state under analysis. Let us thus consider an internal symmetry generated by an operator $Q$ and a pure state $\ket{\Psi}$ which is symmetric under $Q$, namely
\be
[\ket{\Psi}\bra{\Psi},Q]=0.
\ee
It is possible to show that also the reduced density matrix $\rho_A$ is symmetric, meaning that
\be
[\rho_A,Q_A] = 0.
\label{GSsym}
\ee
The latter equation implies that $\rho_A$ is block-diagonal for each block corresponding to different eigenvalues of $Q_A$. An important consequence is the possibility to decompose the R\'enyi and von Neumann entropies according to the symmetry \cite{gs-18}. After introducing the symmetry resolved partition function
\be
\mathcal{Z}_n(q)\equiv \text{tr}(\rho_A^n\Pi_q),
\label{Zn_q}
\ee 
with $\Pi_q$ the projector onto the sector corresponding to the eigenvalue $q$ of $Q_A$\footnote{Note that at this stage this $q$ has nothing to do with the $q$ in the $q$-state Potts model. Unfortunately both uses are standard in the literature so we have decided to keep the standard notation.}, we define the symmetry resolved R\'enyi and von Neumann entropies as follows
\be
S_n(q) \equiv \frac{1}{1-n}\log \frac{\mathcal{Z}_n(q)}{\mathcal{Z}_1(q)^n}, \quad \text{ and } S(q) \equiv -\frac{d}{dn} \left[ \frac{\mathcal{Z}_n(q)}{\mathcal{Z}_1(q)^n}\right]_{n=1}.
\ee
The calculation of all these symmetry resolved quantities requires in general the simultaneous diagonalisation of $\rho_A$ and $Q_A$. An ingenious way to circumvent this difficult path passes through the charged moments
\be
Z_n(\alpha) \equiv \trace\l \rho_A^n e^{i\alpha Q}\r,
\label{cmoments}
\ee
related to $\mathcal{Z}_n(q)$ by Fourier transform. From the definition it is evident that $Z_n(\alpha)=Z_n(\alpha+2\pi)$, meaning that $\alpha$ is identified up to $2\pi$. For continuous $U(1)$ symmetries the charge $q$ is a generic integer, while for discrete $\mathbb{Z}_N$ symmetries $q$ can takes only values $q=0,\dots,N-1$. The relation between the charged moments and the symmetry resolved partition functions reads as follows
\be
\mathcal{Z}_n(q)  = \text{tr}(\rho_A^n\Pi_q) = \begin{cases}\int\limits^{\pi}_{-\pi} \frac{d\alpha}{2\pi}Z_n(\alpha) e^{-i\alpha q} \quad U(1) \text{ case },\\
\frac{1}{N}\sum\limits^{N-1}_{k=0} Z_n\l \frac{2\pi k}{N}\r e^{-i \frac{2\pi k q}{N}} \quad \mathbb{Z}_N \text{ case }\,,\end{cases}
\ee
that is either a continuous Fourier transform (for continuous symmetries) or a discrete Fourier transform (for discrete symmetries). 

The notion of symmetry resolved entanglement, that is, the idea that a certain ``amount" of entanglement can be attributed to individual symmetry sectors of a quantum theory, was introduced in \cite{lr-14} and has since been studied for many systems through the computation of symmetry resolved entropies (SREs). They have been studied for 1+1 conformal field theories (CFTs)
\cite{lr-14,gs-18,Equipartitioning,SREQuench,bc-20,crc-20,Zn,mbc-21, Chen-21, cc-21, cdmWZW-21},
free \cite{mdc-20b,U(1)FreeFF} and interacting integrable quantum field theories (QFTs) \cite{Z2IsingShg, SGSRE},
holographic settings \cite{znm,znm2}, lattice models
\cite{lr-14,Equipartitioning,SREQuench,brc-19,fg-20,FreeF1,FreeF2,mdc-20,ccdm-20,pbc-20,bc-20,HigherDimFermions,mrc-20}, out-of-equilibrium situations \cite{SREQuench, pbc-20,ncv-21,fg-21, pbc-21}
and for other systems exhibiting more exotic types of dynamics \cite{trac-20,MBL,MBL2,Topology,Anyons, as-20}.
Notably, symmetry resolved quantities can be measured experimentally
\cite{ncv-21,vecd-20} which provides a leading motivation for their investigation. 

In a path integral approach to  QFTs, the computation of the charged moments $\trace\l \rho_A^n e^{i\alpha Q}\r$ relies on the evaluation of the partition function of an $n$-copy QFT with specific boundary conditions for the fundamental field $\phi_j$ associated to the $j$-th copy. For 1+1 relativistic QFT, the latter partition function can be equivalently expressed as an expectation value of product of fields, called branch point twist fields (BPTFs), which directly implement the different boundary conditions. At criticality one can use specific techniques of CFT to deal with these correlation functions, providing exact results in different situations \cite{cc-04}. Away from criticality, the exact determination of the correlation functions is instead known to be an extremely difficult task, except for the case of free theories. Nevertheless, the form factor bootstrap program is a powerful tool to investigate IQFTs systematically via the computation of the the form factors, namely the amplitudes of local operators between the vacuum and the multiparticle states\cite{SmirnovBook,KarowskiU1}. Although all these amplitudes are in principle computable, the multi-point correlation functions at large distances are generically dominated by the first few (lower-particle) form factors. For this reason this technique applies efficiently for the infrared properties of these theories. For BPTFs this was first shown in \cite{Ola}.
 
Previous works  involving some of the present authors have focused on theories with $\mathbb{Z}_2$ symmetry\cite{Z2IsingShg}, such as the Ising and sinh-Gordon models, and theories with $U(1)$ symmetry, as is the case of complex free bosons and fermions \cite{U(1)FreeFF} and sine-Gordon model \cite{SGSRE}. In the current work we take a step further by considering a more complex discrete symmetry, that is $\mathbb{Z}_3$, which is present in the $3$-state Potts model. The choice of the model is motivated by the fact that, a part from the Ising model, it is the simplest model displaying an abelian $\mathbb{Z}_N$ symmetry; thus it is natural to consider it as a starting point for the investigation of  integrable perturbations of $\mathbb{Z}_N$ parafermionic CFT's \cite{fz-91}.
Thus, the bulk of this paper is devoted to investigating the explicit construction of form factors via bootstrap techniques for this model. In doing so, our analysis generalizes what has been already done in Ref. \cite{Zn}, where the authors considered the critical version of this model and compared CFT predictions with lattice results. The same internal $\mathbb{Z}_3$ symmetry present in critical ground state ensures that Eq. \ref{GSsym} is satisfied for the off-critical theory. 
 
As mentioned earlier, the SRE can be related to correlators of symmetry resolved BPTFs whereas standard entanglement measures involve BPTFs. Both types of field are associated to symmetries, the former to the combined internal (in our case $\mathbb{Z}_3$) and cyclic permutation symmetry of the theory, the latter only to cyclic permutation symmetry among replicas. For the $q$-state Potts model and any values of $q$, the form factors of neither field have been studied,  thus we devote a small part of our work to considering also the model with $q\leq 3$ and its (total) entanglement entropy. 

The analysis of three-state Potts has some similarities with earlier studies of free theories with $U(1)$ symmetry, since in both cases just two species of particles related by charge conjugation are present; the additional technical difficulty in the present case is the presence of a non-trivial scattering matrix which makes the theory truly interacting. Our main result is the computation of the two-particle form factors, which capture the leading features of all entanglement measures for large connected subsystems.

\medskip

This paper is organized as follows: In Section 2 we review the scattering theory of the $q$-state Potts model and focus particularly on the case $q=3$. In Section 2 we discuss the various types of twist field that are present in the $n$-replica 3-state Potts model: the $\mathbb{Z}_3$ twist fields, the cyclic permutation symmetry branch point twist fields $\TT_n, \tilde{\TT}_n$ and the composite twist fields $\TT_n^\pm, \tilde{\TT}_n^\pm$. For all these fields we discuss their equal-time exchange relations with local fields and their form factor equations. We solve these equations for one and two particles and test our solutions against the $\Delta$-sum rule. In Section 4 we employ our form factor solutions for the composite twist fields to compute their two point function, which is related to the charged moments. These are the building blocks of the symmetry resolved R\'enyi and von Neumann entropies. We compute the leading and next-to-leading large distance correction to the moments and compute the associated entropies. We analyze in detail the analytic continuation of all quantities in the replica number $n$.  We conclude in Section 5. We have also included several appendices. In Appendix A we review some results for the form factors of branch point twist fields in the regime $q\leq 3$ and compute the associated entanglement entropy. We consider all quantities in the ordered phase of the model, where the spectrum is described in terms of excitations called kinks.  We show that in this regime the form factors can be obtained as continuous functions of $q$ and test their validity against the $\Delta$ sum rule. We also show that the next-to-leading order large distance correction to the entropy is proportional to the number of kinks in the model, $q-1$. In Appendix B we review the properties of the minimal form factor in the disordered phase. In Appendix C we analyze the asymptotics of the R\'enyi entropy and symmetry resolved R\'enyi entropy for $n\rightarrow \infty$.

\section{The $q$-State Potts Model: Scattering Theory}
The scattering matrix of the $q$-state Potts model was first computed in \cite{CZam} for $0<q<4$. For $q=3$ the scattering theory is particularly simple and the associated $S$-matrices were studied in \cite{SaJiMi,ZaPotts}. On the other hand, for $q>4$ the scattering theory becomes increasingly complicated, particularly the pole structure of the scattering amplitudes so that demonstrating closure of the bootstrap equations is highly non-trivial \cite{DoPoTa1, DoPoTa2}.

In this section we focus on the scattering theory and particle content of the model, closely following the presentation in \cite{Potts_Delf}.  In the $q$-state Potts model, physical states are constructed in terms of ``kink'' operators $K_{\alpha \beta}(\theta)$ which represent kinks interpolating between vacua $\alpha, \beta$. The allowed vacua are labeled by indices $\alpha=1,2,\ldots, q$. From this definition it is natural to think of $q$ as an integer. However, it is possible to make sense of the model for generic $q$ by appealing to its original definition as a lattice theory. In general, a $p$-kink state may be written as 
\beq 
|K_{\alpha_0 \alpha_1}(\theta_1) K_{\alpha_1\alpha_2}(\theta_2) \ldots K_{\alpha_{p-1}\alpha_p}(\theta_p)\rangle \quad \text{with}\quad \alpha_i \neq \alpha_{i+1} \quad \forall \, i=1,\ldots, p,
\eeq 
where $\theta_i$ are the rapidity variables. 
Such a state is ``neutral'' if $\alpha_0=\alpha_p$ and ``charged" if $\alpha_0 \neq \alpha_p$.  Since the model is integrable the scattering theory is fully determined by the two-particle scattering amplitudes. In addition the permutation symmetry under exchange of the vacua implies that there are only four independent functions that need to be computed namely the Zamolodchikov-Faddeev (ZF) \cite{ZZ, FA} algebra may be written as
\beqa 
K_{\alpha \gamma}(\theta_1)K_{\gamma \beta}(\theta_2)&=&S_0(\theta_{12})
\sum_{\delta \neq \gamma} K_{\alpha \delta}(\theta_2)K_{\delta \beta}(\theta_1) + S_1(\theta_{12}) K_{\alpha \gamma} (\theta_2) K_{\gamma \beta}(\theta_1) \quad \alpha \neq \beta, \nonumber\\
K_{\alpha \gamma}(\theta_1)K_{\gamma \alpha}(\theta_2)&=&S_2(\theta_{12})
\sum_{\delta \neq \gamma} K_{\alpha \delta}(\theta_2)K_{\delta \alpha}(\theta_1) + S_3(\theta_{12}) K_{\alpha \gamma} (\theta_2) K_{\gamma \alpha}(\theta_1).
\eeqa 
where, as usual, $\theta_{ij}=\theta_i-\theta_j$ and the first term in the relations above indicates that the scattering is generally non-diagonal. 
The amplitudes $S_i(\theta)$ with $i=0,1,2,3$ are constrained by a number of equations related to physical requirements such as unitarity. These equations can be solved analytically giving 
\beqa 
S_0(\theta)&=&\frac{\sinh \lambda \theta \sinh \lambda(\theta-i\pi)}{\sinh\lambda \left(\theta-\frac{2\pi i}{3} \right)\sinh\lambda \left(\theta-\frac{i\pi}{3} \right)}S(\theta),\nonumber\\
S_1(\theta)&=& \frac{\sin\frac{2\pi\lambda}{3} \sinh \lambda(\theta-i\pi)}{\sin \frac{\pi \lambda}{3} \sinh\lambda \left(\theta-\frac{2\pi i}{3} \right)}S(\theta),\nonumber\\
S_2(\theta)&=& \frac{\sin\frac{2\pi\lambda}{3} \sinh \lambda\theta}{\sin \frac{\pi \lambda}{3} \sinh\lambda \left(\theta-\frac{i\pi}{3} \right)}S(\theta),\nonumber\\
S_3(\theta)&=& \frac{\sin\lambda \pi}{\sin \frac{\pi \lambda}{3}}S(\theta),
\eeqa 
in terms of the variable $\lambda$  which is related to $q$ as follows,
\begin{equation} 
\sqrt{q}=2 \sin\frac{\pi\lambda}{3}\,.
\label{lam}
\end{equation}
From this definition it follows that $q$ is only integer for very particular values of $\lambda$. For $\lambda=\frac{3}{2}$ we have $q=4$ and the resulting theory has 4 particles and $S$-matrices which can be identified with those of the $D_4$-minimal Toda theory. Similarly $\lambda=1$ corresponds to $q=3$, $\lambda=9/4$ to $q=2$ and $\lambda=5/2$ to $q=1$. 

The function $S(\theta)$ may be expressed as an infinite product of gamma functions or also through an integral representation given by
\beq 
S(\theta)=\frac{\sinh \lambda\left(\theta-\frac{i\pi}{3} \right)}{\sinh \lambda\left(\theta-i\pi\right)} e^{\mathcal{A}(\theta)},
\eeq 
with 
\beq 
\mathcal{A}(\theta)=\int_{0}^\infty \frac{dt}{t} 
\frac{\sinh \frac{t}{2}\left(1-\frac{1}{\lambda}\right)-\sinh \frac{t}{2}\left(\frac{1}{\lambda}-\frac{5}{3}\right)}{\sinh \frac{t}{2\lambda} \cosh \frac{t}{2}} \sinh \frac{t\theta}{i\pi}.
\eeq 
For $\lambda\leq 1$ the function $S(\theta)$ has no poles on the physical sheet. However, the amplitudes $S_0(\theta)$ and $S_1(\theta)$ have a pole at $\theta=\frac{2\pi i}{3}$ in the direct channel corresponding to the formation of a bound state, which is itself an elementary kink $K$ with three point coupling
\beq 
(\Gamma_{KK}^K)^2={\frac{1}{\lambda} \sinh \frac{2\pi \lambda}{3} e^{\mathcal{A}(\frac{i\pi}{3})}}.
\eeq 
Correspondingly, the amplitudes $S_2(\theta)$ and $S_3(\theta)$ have a pole in the cross-channel at $\theta=\frac{i\pi}{3}$. 

For $\lambda>1$ an additional pole in the direct channel of the amplitudes $S_2(\theta)$ and $S_3(\theta)$ enters the physical strip at 
\beq 
\theta_B=i\pi \left(1-\frac{1}{\lambda}\right).
\label{thetaB}
\eeq 
This is associated with the formation of a kink-kink bound state $B$ of mass
$m_B=2m \cosh\frac{\theta_B}{2}$ and three-point coupling
\beq 
(\Gamma_{KK}^B)^2=\frac{1}{\lambda} \sin \frac{4\pi \lambda}{3} \frac{\sin \pi \lambda}{\sin \frac{\pi \lambda}{3}} e^{\mathcal{A}(\theta_B)}.
\eeq 
This bound state is a new particle with its own scattering matrix elements. These can be expressed in terms of fundamental blocks:
\beq 
(\theta)_a=\frac{\tanh\frac{1}{2}\left(\theta + i \pi a \right)}{\tanh\frac{1}{2}\left(\theta - i \pi a \right)}\,,
\eeq 
as
\beqa 
S_{KB}(\theta)&=&(\theta)_{1-\frac{\theta_B}{2\pi i}}(\theta)_{\frac{2}{3}-\frac{\theta_B}{2\pi i}},\\
S_{BB}(\theta)&=& (\theta)_{\frac{2}{3}}(\theta)_{1-\frac{\theta_B}{\pi i}}(\theta)_{\frac{2}{3}-\frac{\theta_B}{\pi i}}.
\eeqa 
For $q\leq 4$ or $0\leq \lambda \leq \frac{3}{2}$ this describes the full particle spectrum of the theory. 
For  $q=3$, the $S$-matrices reduce simply to \cite{CZam,SaJiMi,ZaPotts}
\beq
S_1(\theta)= \frac{\sinh\frac{1}{2}\left(\theta+\frac{2\pi i}{3}\right)}{\sinh\frac{1}{2}\left(\theta-\frac{2\pi i}{3}\right)}\,\quad S_2(\theta)= -\frac{\sinh\frac{1}{2}\left(\theta+\frac{\pi i}{3}\right)}{\sinh\frac{1}{2}\left(\theta-\frac{\pi i}{3}\right)}\quad \mathrm{and}\quad S_3(\theta)=0\,. 
\label{3q}
\eeq

\subsection{3-State Potts Model}\label{3Potts}
Since much of this paper is devoted to the $3$-state Potts model, let us discuss this case in a bit more detail. 
The model can be seen as a perturbed CFT \cite{Potts_Delf}, with Euclidean action given by
\be
S = S_{\text{CFT}} + \tau \int d^2x \, \varepsilon(\texttt{x})\,.
\label{mass}
\ee
The underlying conformal field theory \cite{Zn_Zam2}, obtained for $\tau=0$, is the tricritical $3$-state Potts model, that is the minimal model $\mathcal{M}(6,5)$, of central charge $c= 4/5$,  and the energy density field $\varepsilon(x)$ is identified with the primary field $\Phi_{2,1}$ with conformal dimensions
\be
(\Delta_\varepsilon,\bar{\Delta}_{\varepsilon}) = \l \frac{2}{5}, \frac{2}{5}\r\,.
\ee
This CFT is also called parafermionic $\mathbb{Z}_3$ theory and it was studied in detail in \cite{Zn_Zam2}. It is possible to show that this model is integrable and, depending on the sign of $\tau$, it develops a paramagnetic or ferromagnetic phase. In our study of the $3$-state Potts model we have focused on the paramagnetic (or disordered) phase (see \cite{Part_Delfino,Zn_integ}), in contrast with the analysis of the ferromagnetic phase (or ordered) of \cite{Potts_Delf} which is the point of view described at the beginning of this Section for more general $q$. However, the choice of phase for $q=3$ makes little difference, at least in what concerns the scattering theory of the model. Instead it amounts to a change in the way we identify the fundamental particles of the theory. Typically, in the paramagnetic phase we consider the particles as fundamental excitations above a vacuum $\ket{0}$ instead of the kinks interpolating different vacua. Multiparticle states are then constructed in terms of particles belonging to one of two species, say $A$ and $\bar{A}$. The scattering is diagonal and the only non-trivial scattering phases are the transmission amplitudes
\be
S_{AA}(\theta) \equiv S^{AA}_{AA}(\theta) = \frac{\sinh\frac{1}{2} (\theta+\frac{2\pi i}{3})}{\sinh\frac{1}{2} (\theta-\frac{2\pi i}{3})},\quad S_{A\bar{A}}(\theta) \equiv S^{A\bar{A}}_{A\bar{A}}(\theta)  = -\frac{\sinh\frac{1}{2}(\theta+\frac{i\pi}{3})}{\sinh\frac{1}{2}(\theta-\frac{i\pi}{3})},
\ee
where $\theta$ is the rapidity difference among the two particles. The latter amplitudes are related by crossing symmetry $S_{AA}(\theta) = S_{A\bar{A}}(i\pi-\theta)$. Moreover, since the theory is invariant under charge conjugation and parity symmetry, then $S_{\bar{A}\bar{A}}(\theta) = S_{AA}(\theta)$ and $S_{A\bar{A}}(\theta) = S_{\bar{A}A}(\theta)$. Note that these amplitudes are identical to (\ref{3q}).
\begin{figure}[t]
\centering
	\includegraphics[width=0.48\linewidth]{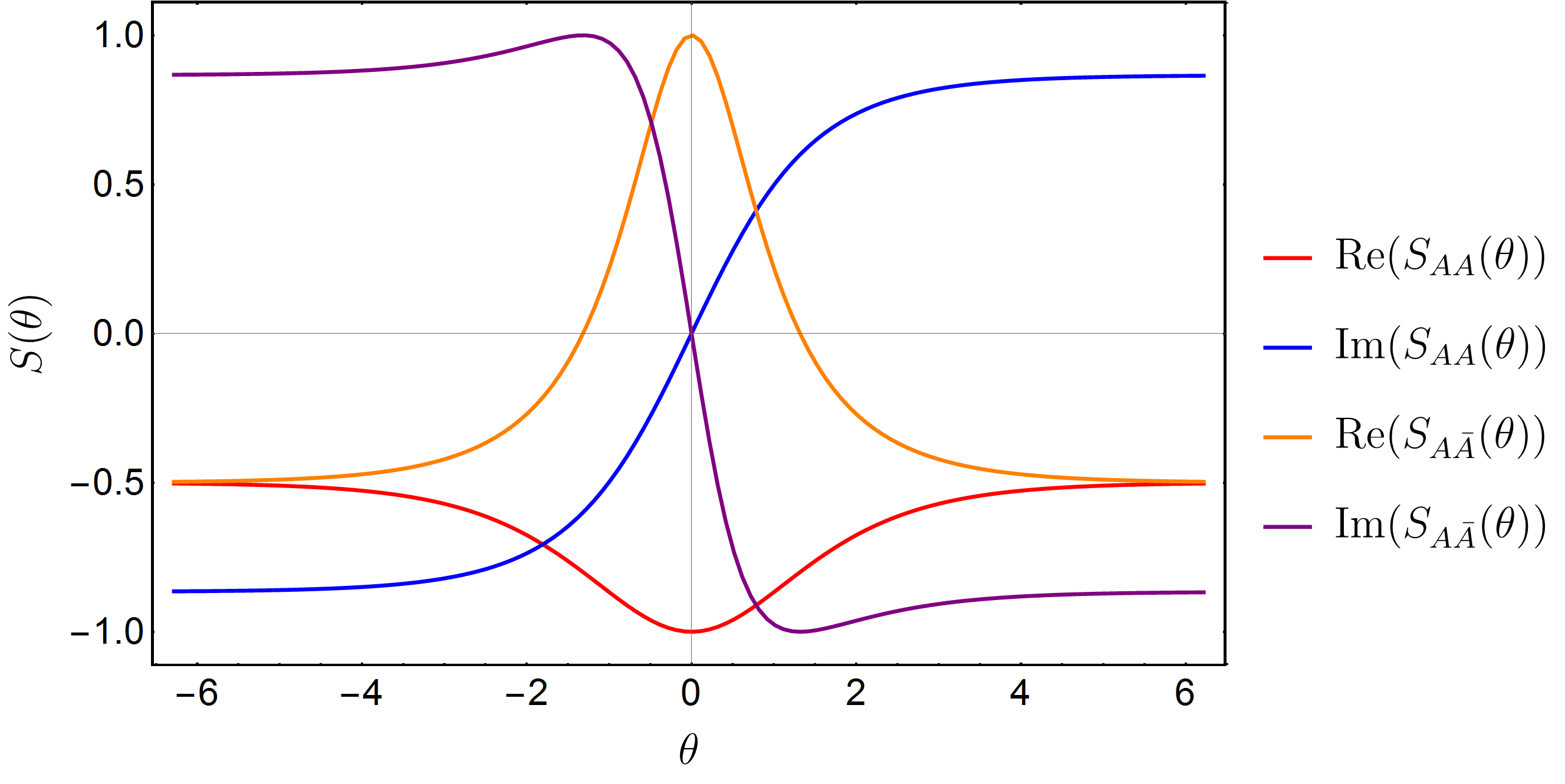}
	\includegraphics[width=0.48\linewidth]{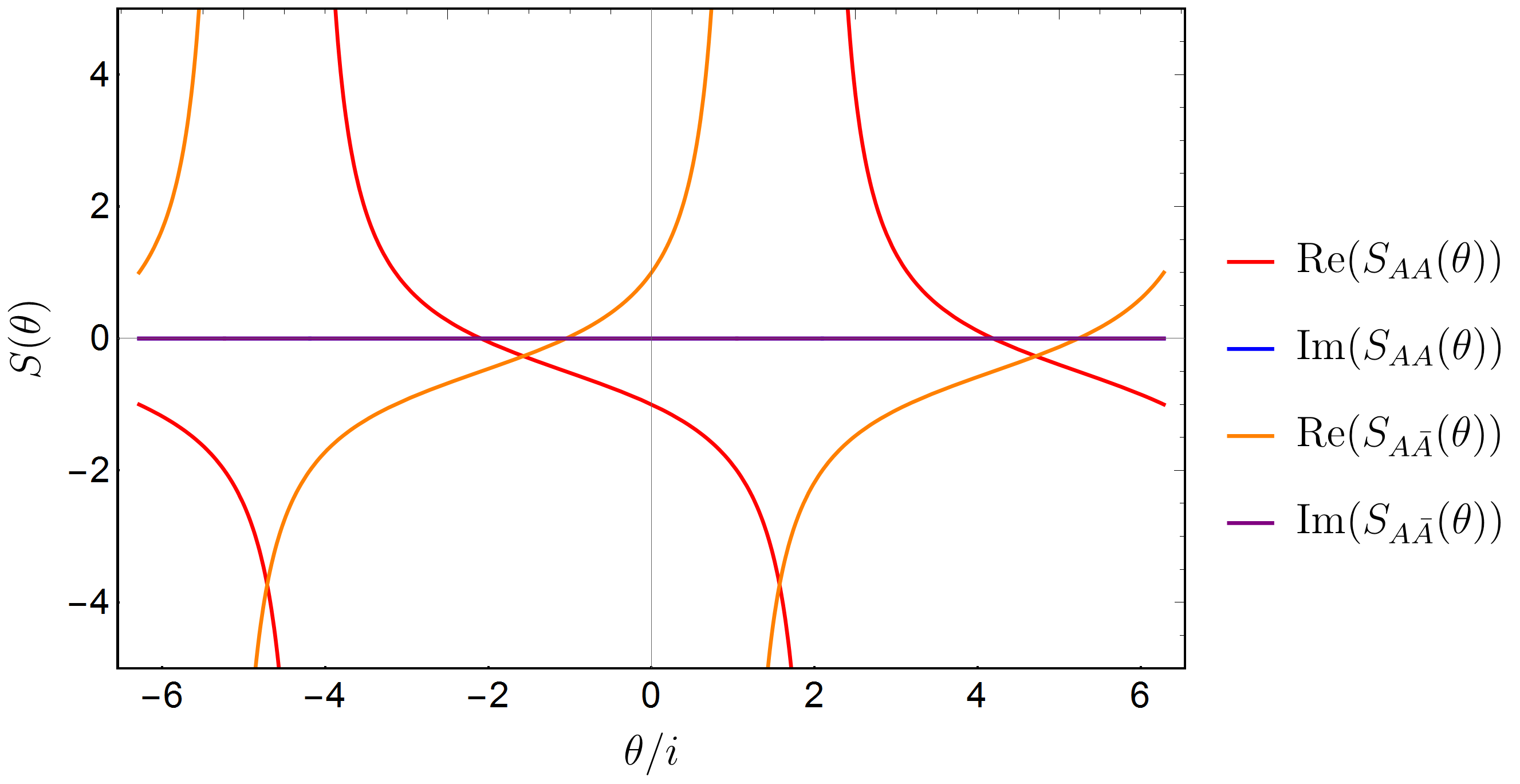}
	\caption{Left: Scattering amplitude for real rapidity difference ($\text{Im}(\theta)=0$) in the 3-state Potts model. When $\theta=0$ we have $S_{A\bar{A}}(\theta=0) =1$ and $S_{AA}(\theta=0) =-1$, while for $|\theta|\rightarrow +\infty$ the scattering amplitude goes to a constant. Right: Scattering amplitude for imaginary values of the rapidity difference ($\text{Re}(\theta)=0$) in the 3-state Potts model. The imaginary part is vanishing and the amplitudes develop dynamical poles associated with the presence of a bound state. }
	\label{Smatrix_Im}
\end{figure}
A dynamical pole is present for $S_{AA}(\theta)$ with associated residue
\be
\underset{\theta = i2\pi/3}{\text{Res}}S_{AA}(\theta) =(\Gamma_{AA}^{\bar{A}})^2= i\sqrt{3}\,,
\ee
which corresponds to the bound state formation process
\be
A+ A\rightarrow \bar{A}\,.
\label{process}
\ee
In Fig. \ref{Smatrix_Im} we plot the S-matrix as a function of the rapidity difference $\theta$ along the real and the imaginary axis respectively. An important feature is that $S_{AA}(0) =-1$ which implies fermionic statistics; for this reason we call $S_{AA}(\theta)$ a fermionic-type scattering matrix. Instead $S_{A\bar{A}}(0) =1$ and similarly $S_{A\bar{A}}(\theta)$ are bosonic-type scattering matrices.

\medskip
Critical to our analysis of the symmetry resolved entanglement is the fact that 
an internal $\mathbb{Z}_3$ symmetry is present for this theory. It is generated by the following action on the particles
\be
A \rightarrow e^{\frac{2\pi i}{3}}A, \qquad  \bar{A} \rightarrow e^{-\frac{2\pi i}{3}}\bar{A}.
\ee
We can then associate a $\mathbb{Z}_3$-charge to each particle, which is additive and equals $+1$ for species $A$, and $-1$ for species $\bar{A}$. Note that the charge is conserved in the process (\ref{process}), since
\be
(+1)+(+1) = -1 \quad (\text{mod } 3).
\ee
Similar to the Ising model, in the 3-state Potts model we also find order and disorder fields which are either local or semi-local w.r.t. the  $\mathbb{Z}_3$ symmetry. We call them $\sigma_1(\texttt{x}),\mu_1(\texttt{x})$ respectively. While $\sigma_1({\bf x})$ is local w.r.t. the particles and has the meaning of a magnetization operator, $\mu_1(\texttt{x})$ is non-local and acts nontrivially introducing an Aharonov-Bohm flux of value $e^{i2\pi/3}$ for each space point $y^1>x^1$. In other words, $\mu_1({\bf x})$ is the twist field associated with $\mathbb{Z}_3$ symmetry. The conjugate fields are $\sigma_{-1}(\texttt{x})$  and  $ \mu_{-1}(\texttt{x})$ and the latter gives an Aharonov-Bohm flux of value $e^{-i2\pi/3}$. The two order/disorder operators $\mu_{\pm 1},\sigma_{\pm  1}$ share the same conformal dimensions
\be
(\Delta_{\sigma_{\pm 1}},\bar{\Delta}_{\sigma_{\pm 1}}) =(\Delta_{\mu_{\pm 1}},\bar{\Delta}_{\mu_{\pm 1}}) = \l \frac{1}{15}, \frac{1}{15} \r,
\ee
which dictate the ultraviolet behaviour of their correlation functions. At criticality, these fields are identified with the primary field $\Phi_{2,3}$ of the minimal model $\mathcal{M}(6,5)$. Away from criticality, one can relate the paramagnetic and ferromagnetic phase via the Kramers-Wannier duality, which exchanges the role of the order and disorder operators. This last observation is useful as it implies that we can compare results for the disorder operator in the paramagnetic phase to those for the order operator in the ferromagnetic phase, as discussed in \cite{Potts_Delf}.




\section{Twist Fields in the Potts Model}
Twist fields play an important role in QFT and they are present in conjunction with internal symmetries of the theory. The best known examples are the $\sigma/\mu$ order/disorder fields in the Ising model or, as we have just discussed, the fields $\sigma_{\pm 1}/\mu_{\pm 1}$ in the 3-state Potts model; these fields are associated with  $\mathbb{Z}_2$ or $\mathbb{Z}_3$ symmetry respectively.  Twist fields are typically non-local or semi-local with respect to other quantum fields in the theory. This non-locality translates into non-trivial equal-time exchange relations between twist fields and local fields, which can be easily proven by making twist fields sit at branch points in space-time, that is at the origin of branch cuts. For our purposes three different types of twist field have to be considered:

\begin{itemize}
\item[1)] the disorder fields $\mu_{\pm 1}$ associated with $\mathbb{Z}_3$ symmetry of the 3-state Potts model in its disordered phase, 
\item[2)] the BPTFs, $\mathcal{T}_n$ and its conjugate $\tilde{\mathcal{T}}_n$, associated with cyclic permutation symmetry among copies in the $n$-replica $q$-state Potts model, which play a central role in computations of the entanglement entropy,
\item[3)] for $q=3$ the ``fusion" of these two classes of fields, as originally described in \cite{cdl-12,leviFFandVEV,bcd-15} gives rise to composite twist fields (CTFs) $\mathcal{T}^{\pm}_n$ and $\tilde{\mathcal{T}}^{\pm}_n$. These fused fields
play a central role in the computation of the symmetry resolved entanglement entropies, as described in \cite{gs-18}.  
 \end{itemize}
The exchange relations for standard BPTFs in a replica QFT can be written as \cite{Ola}
\begin{eqnarray}
\mathcal{O}_{i}({\bf y})\mathcal{T}_n({\bf x}) & = & \mathcal{T}_n({\bf x})\mathcal{O}_{i+1}({\bf y})\quad\mathrm{for}\quad y^{1}>x^{1}\,,\label{BPTFSpatialExchange}\\
 & = & \mathcal{T}_n({\bf x})\mathcal{O}_{i}({\bf y}) \qquad \mathrm{for}\quad x^{1}>y^{1}\,,
\end{eqnarray}
with respect to quantum fields $\mathcal{O}_{i}$ living on the
$i$-th replica.
Similarly, if $\tilde{\mathcal{T}_n}({\bf x})$ is the hermitian conjugate of ${\mathcal{T}_n}({\bf x})$ it is then associated with the inverse cyclic permutation, and we can also define $\tilde{\mathcal{T}_n}$ with exchange relations
\begin{eqnarray}
\mathcal{O}_{i}({\bf y})\tilde{\mathcal{T}_n}({\bf x}) & = & \tilde{\mathcal{T}_n}({\bf x})\mathcal{O}_{i-1}({\bf y})\quad\mathrm{for}\quad y^{1}>x^{1}\,,\label{BPTFSpatialExchange(AntiTwistField)}\\
 & = & \tilde{\mathcal{T}_n}({\bf x})\mathcal{O}_{i}({\bf y})\qquad   \mathrm{for}\quad x^{1}>y^{1}\,.
\end{eqnarray}
Let us now consider the local operator $\mathcal{V}_\alpha(x)$ which introduces an Aharonov-Bohm flux in the region $y^1>x^1$.  For the moment,  one can even assume that $\alpha$ is a continuous parameter for simplicity. Doing so, the generator of the symmetry $e^{i\alpha Q_{A}}$ of the subsystem $A=[0,\ell]$ can be identified by
\be
e^{i\alpha Q_{A}} \sim \mathcal{V}_\alpha(0)(\mathcal{V}_\alpha)^\dagger(\ell).
\ee
The mutual locality between $\mathcal{V}_\alpha$ and an another operator $\mathcal{O}$ is identified by the relation
\begin{equation}
\mathcal{O}(y,t')\mathcal{V}_{\alpha}(x,t)=e^{i\kappa_{\mathcal{O}}\alpha}\mathcal{V}_{\alpha}(x,t)\mathcal{O}(y,t'),
\label{eq:LocalityDef}
\end{equation}
or, when using the radial quantisation picture, 
\begin{equation}
\mathcal{O}(0,0)\mathcal{V}_{\alpha}(e^{-i2\pi}z,e^{i2\pi}\bar{z})=e^{i\kappa_\mathcal{O}\alpha}\mathcal{O}(0,0)\mathcal{V}_{\alpha}(z,\bar{z}).
\end{equation}
We refer to $\kappa_\mathcal{O}$ as the charge of the operator $\mathcal{O}$ for the symmetries under consideration. The fusion between the standard twist field $\mathcal{T}_n$ and $\mathcal{V}_\alpha$ gives rise to the so-called composite twist fields which can be defined very precisely in CFT  \cite{cdl-12, leviFFandVEV, bcd-15}
\begin{equation}
:\!\mathcal{T}_n\,\mathcal{V}_\alpha:(y)\!:= n^{2\Delta_{\alpha} -1}\lim_{x\rightarrow y} |x-y|^{2 \Delta_\alpha (1-\frac{1}{n})} \sum_{j=1}^n \mathcal{T}_n(y) \mathcal{V}_{\alpha,j}(x)\,,
\label{ccft}
\end{equation}
 where $\mathcal{V}_{\alpha,j}(x)$ is the copy of field $\mathcal{V}_{\alpha}(x)$ living in replica $j$, $:\bullet:$ represents normal ordering and the power law involves the conformal dimension $\Delta_\alpha$ of the field $\mathcal{V}_\alpha$ (for spinless fields, this is half of the scaling dimension $h_\alpha$ given earlier).
 
For the $\mathbb{Z}_3$-symmetry of the Potts model we just need to identify $\mathcal{V}_\alpha$ for the values $\alpha=0,\pm 2\pi/3$. $\mathcal{V}_{0}$ is the identity operator by definition, while, as pointed out in section \ref{3Potts}, for the other values of the flux one has
\be
\mathcal{V}_{\pm2\pi/3} = \mu_{\pm 1}.
\ee
We then define the fields $\TT_n^\tau(\mathbf{x})$, $\tilde{\TT}_n^{\tau}(\mathbf{x})$ with $\tau=\pm,0$ as
\begin{eqnarray}
\mathcal{T}_n^\pm(\mathbf{x})&:=&\, :\!\mathcal{T}_n\,\mu_{\pm 1}\!:(\mathbf{x})\, \qquad \mathrm{and} \qquad \mathcal{T}_n^0(\mathbf{x}):=\mathcal{T}_n(\mathbf{x})\,, \nonumber\\
\tilde{\mathcal{T}}_n^\mp(\mathbf{x})&:=&\, :\!\tilde{\mathcal{T}}_n\,\mu_{\pm 1}\!:(\mathbf{x})\, \qquad \mathrm{and} \qquad \tilde{\mathcal{T}}_n^0(\mathbf{x}):=\tilde{\mathcal{T}}_n(\mathbf{x})\,.
\label{ccft2}
\end{eqnarray}
The definition (\ref{ccft}) is valid in the UV CFT in a strict sense, but the off-critical version of the composite operator can as usual be interpreted as the operator (\ref{ccft}) flows to under the massive perturbation (\ref{mass}). In particular, the off-critical fields of interest in the replica Potts model are characterised by equal time exchange relations 
\begin{eqnarray}
\mathcal{O}_{i}({\bf y})\mathcal{T}_n^{\tau}({\bf x}) & = & e^{\frac{i\kappa \tau 2\pi}{3}}\mathcal{T}_n^{\tau}({\bf x})\mathcal{O}_{i+1}({\bf y})\quad\mathrm{for}\quad y^{1}>x^{1} \text{ and } i=n\,,\label{CTFSpatialExchange}\\
 & = & \mathcal{T}_n^{\tau}({\bf x})\mathcal{O}_{i}({\bf y})\qquad \quad\quad\quad \,\, \,\mathrm{otherwise}\,,
\end{eqnarray}
for any quantum field $\mathcal{O}_{i}$ living on the $i$th replica and possessing a definite $\mathbb{Z}_3$ charge $\kappa=\pm1,0$.
Similarly, as $\tilde{\mathcal{T}_n}({\bf x})$ is the hermitian conjugate of ${\mathcal{T}_n}({\bf x})$ associated with the inverse cyclic permutation, we can also define $\tilde{\mathcal{T}_n}^{\tau}$ with exchange relations
\begin{eqnarray}
\mathcal{O}_{i}({\bf y})\tilde{\mathcal{T}_n}^{\tau}({\bf x}) & = & e^{-\frac{i\kappa_\mathcal{O} \tau 2\pi}{3}}\tilde{\mathcal{T}_n}^{\tau}({\bf x})\mathcal{O}_{i-1}({\bf y})\quad\mathrm{for}\quad y^{1}>x^{1} \text{ and } i=n\,,\label{CTFSpatialExchange(AntiTwistField)}\\
 & = & \tilde{\mathcal{T}_n}^{\tau}({\bf x})\mathcal{O}_{i}({\bf y})\qquad \quad\quad\quad \,\,\,\,\, \,\mathrm{otherwise}\,.
\end{eqnarray}
Note that the exchange relations for $\TT_n^\tau$ reduce to those for $\TT_n$ when $\tau=0$, in accordance with the identifications in (\ref{ccft2}).

\subsection{Form Factors of Branch Point Twist Fields\label{BPTF}}

With the exchange relations (\ref{BPTFSpatialExchange}) at hand, one can formulate BPTF form factor equations in integrable QFTs (IQFTs), which generalize the standard form factor programme for local fields \cite{SmirnovBook,KarowskiU1}.
These equations were first given in \cite{Ola} for diagonal theories
and then in \cite{cd-08} for non-diagonal ones. Although the FF  bootstrap equations for the Potts model have been studied in \cite{Potts_Delf}, to the best of our knowledge these studies focuss on the ordered phase of the model. Our main goal here is to study the symmetry resolved fields or CTFs in the disordered phase. To this end it is well justified to specify the bootstrap equations for the standard BPTF in the disordered phase and briefly discuss their solutions as well. The FFs of the BPTF in the ordered phase can be constructed very much along the lines of \cite{Potts_Delf} and 
a computation for $q\leq 3$ is presented in Appendix \ref{FFBPTF}.  These equations and their solutions are an important reference point for our later investigations as they allow us to conveniently introduce  some elements of IQFT as well as our notations.

The most important objects are the form factors (FF), which are matrix elements of (semi-)local operators
$\mathcal{O}(x,t)$ between the vacuum and asymptotic state, i.e., 
\begin{equation}
F_{\gamma_{1}\ldots\gamma_{k}}^{\mathcal{O}}(\theta_{1},\ldots,\theta_{k})=\langle0|\mathcal{O}(0,0)|\theta_{1},\ldots\theta_{k}\rangle_{\gamma_{1}\ldots\gamma_{k}}.\label{eq:FF}
\end{equation}
In massive field theories like the Potts model, the asymptotic states
are spanned by multi-particle excitations whose dispersion relation
can be parametrised as $(E,p)=(m_{\gamma_{i}}\cosh\theta,m_{\gamma_{i}}\sinh\theta)$,
where $\gamma_{i}$ indicates the particle species ($A$ or $\bar{A}$ for us) and $\theta$
is the rapidity of the particle. In such models, any multi-particle
state can be constructed from the vacuum state $|0\rangle$ as 
\begin{equation}
|\theta_{1},\theta_{2},...,\theta_{k}\rangle_{\gamma_{1}\ldots\gamma_{k}}=Z_{\gamma_{1}}^{\dagger}(\theta_{1})Z_{\gamma_{2}}^{\dagger}(\theta_{2})\ldots.Z_{\gamma_{k}}^{\dagger}(\theta_{k})|0\rangle\:,\label{eq:basis}
\end{equation}
where $Z^{\dagger}$s are particle creation operators; in particular
the operator $Z_{\gamma_{i}}^{\dagger}(\theta_{i})$ creates a particle
of species $\gamma_{i}$ with rapidity $\theta_{i}$. In an IQFT with
factorised scattering, the creation and annihilation operators $Z_{\gamma_{i}}^{\dagger}(\theta)$
and $Z_{\gamma_{i}}(\theta)$ satisfy the  ZF
algebra \cite{ZZ,FA} which in the diagonal case reads
\begin{eqnarray}
Z_{\gamma_{i}}^{\dagger}(\theta_{i})Z_{\gamma_{j}}^{\dagger}(\theta_{j}) & = & S_{\gamma_{i}\gamma_{j}}(\theta_{ij})Z_{\gamma_{j}}^{\dagger}(\theta_{j})Z_{\gamma_{i}}^{\dagger}(\theta_{i})\:,\nonumber \\
Z_{\gamma_{i}}(\theta_{i})Z_{\gamma_{j}}(\theta_{j}) & = & S_{\gamma_{i},\gamma_{j}}(\theta_{ij})Z_{\gamma_{j}}(\theta_{j})Z_{\gamma_{i}}(\theta_{i})\:,\nonumber \\
Z_{\gamma_{i}}(\theta_{i})Z_{\gamma_{j}}^{\dagger}(\theta_{j}) & = & S_{\gamma_{i}\gamma_{j}}(\theta_{ji})Z_{\gamma_{j}}^{\dagger}(\theta_{j})Z_{\gamma_{i}}(\theta_{i})+\delta_{\gamma_{i},\gamma_{j}}2\pi\delta(\theta_{i}-\theta_{j}),\label{eq:ZF}
\end{eqnarray}
where $S_{\gamma_{i}\gamma_{j}}(\theta_{ij})$
denotes the two-body S-matrices of the theory as function of rapidity differences. In the $n$-replica IQFT, the above algebra is understood as follows: the scattering between particles in different
and in the same copies is described as
\begin{equation}
\begin{split}S_{(\gamma_{i},\nu_{i})(\gamma_{j},\nu_{j})}(\theta)= &\begin{cases}
S_{\gamma_{i}\gamma_{j}}(\theta) & \nu_{i}=\nu_{j}\\
1 & \nu_{i}\neq\nu_{j}
\end{cases}\end{split}
\end{equation}
where we introduced the replica index $\nu_{i}$ which takes
values from $1$ to $n$ and it is identified up $\nu_i \sim \nu_i +n$. To make our notations easier we introduce
the multi-index, following \cite{Ola}

\begin{equation}
a_{i}=(\gamma_{i},\nu_{i})\,,
\end{equation}
together with

\begin{equation}
\bar{a}_{i}=(\bar{\gamma}_{i},\nu_{i}),\qquad\hat{a}_{i}=(\gamma_{i},\nu_{i}+1),
\end{equation}
where  $\bar{\gamma}_{i}$ denotes the
anti-particle of $\gamma_{i}.$ Denoting the FFs of $\mathcal{T}_n$
 by $F_{\underline{a}}(\underline{\theta},n)$, the bootstrap equations can be formulated as 

\begin{eqnarray}
 &  & F_{\underline{a}}(\underline{\theta},n)=S_{a_{i}a_{i+1}}(\theta_{i,i+1})F_{\ldots a_{i-1}a_{i+1}a_{i}a_{i+2}\ldots}(\ldots\theta_{i+1},\theta_{i},\ldots,n),\label{eq:FFAxiom1}\\
 &  & F_{\underline{a}}(\theta_{1}+2\pi i,\theta_{2},\ldots,\theta_{k},n)=F_{a_{2}a_{3}\ldots a_{k}\hat{a}_{1}}(\theta_{2},\ldots,\theta_{k},\theta_{1},n),\label{eq:FFAxiom2}\\
 &  & -i\underset{\theta_{0}'=\theta_{0}+i\pi}{{\rm Res}}F_{\bar{a}_{0}a_{0}\underline{a}}(\theta_{0}',\theta_{0},\underline{\theta},n)=F_{\underline{a}}(\underline{\theta},n),\label{eq:FFAxiom3}\\
 &  & -i\underset{\theta_{0}'=\theta_{0}+i\pi}{{\rm Res}}F_{\bar{a}_{0}\hat{a}_{0}\underline{a}}(\theta_{0}',\theta_{0},\underline{\theta},n)=-\prod_{l=1}^kS_{\hat{a}_{0}a_{l}}(\theta_{0l})F_{\underline{a}}(\underline{\theta},n),\nonumber \\
 &  & -i\underset{\theta_{0}'=\theta_{0}+i\bar{u}_{\gamma \gamma}^{\bar{\gamma}}}{{\rm Res}}F_{(\gamma,\nu_{0})(\gamma,\nu'_{0})\underline{a}}(\theta_{0}',\theta_{0},\underline{\theta},n)=\delta_{\nu_{0},\nu'_{0}}\Gamma_{\gamma \gamma}^{\bar{\gamma}}F_{(\bar{\gamma},\nu_{0})\underline{a}}(\theta_{0},\underline{\theta},n),\label{eq:FFAxiom4}
\end{eqnarray}
where  $\underline{\theta}$
and $\underline{a}$ are shorthands for $\theta_{1},\theta_{2},...,\theta_{k}$
and $(\gamma_{1},\nu_{1})(\gamma_{2},\nu_{2})....(\gamma_{k},\nu_{k})$
respectively, where $\gamma=A,\bar{A}$ and $\bar{\bar{A}}=A$.

In the 3-state Potts model two particles of type $A$ can form bound state $\bar{A}$ or the other way round and this is encoded in the bound state kinematic equation \eqref{eq:FFAxiom4} which links FFs with total particle numbers $k$ and $k-1$. It is easy to see, however, that the one-particle FFs  of BPTF as well as for the composite twist field are vanishing. The reason is that both fields are neutral w.r.t. $\mathbb{Z}_3$ charge. This implies that only FFs with an equal number of particles and anti-particles mod 3 are non-vanishing and consequently the one-particle FFs are zero.
Relativistic invariance has implications
on FFs of BPTFs according to 
\begin{equation}
F_{\underline{a}}(\theta_{1}+\Lambda,\ldots,\theta_{k}+\Lambda)=e^{\Sigma\Lambda}F_{\underline{a}}(\underline{\theta}),\label{eq:RelInv}
\end{equation}
with the Lorentz spin $\Sigma=0$ since the BPTF is spinless.
It follows then that the two-particle FFs depend only on one rapidity variable, that is the rapidity difference. 
Following Ref. \cite{Ola} it is easy to write down the two-particle FF of the BPTF $\mathcal{T}_n$ for particles in the same copy (say 1), which reads as follows
\begin{equation}
F_{(A,1)(\bar{A},1)}(\theta,n)=\frac{\langle\mathcal{T}_{n}\rangle\sin\frac{\pi}{n}}{2n\sinh\frac{i\pi+\theta}{2n}\sinh\frac{i\pi-\theta}{2n}}\frac{h(\theta,n)}{h(i\pi,n)}\,,\label{eq:F2TwistField}
\end{equation}
where $\langle\mathcal{T}_{n}\rangle$ is the
vacuum expectation value (VEV) of ${\cal T}_n$ and $h(\theta,n)$ is an entire function known as the minimal form factor which we present in Appendix \ref{minimal_FF}. From $F_{(A,1)(\bar{A},1)}(\theta,n)$ we obtain $F_{(A,j)(\bar{A},k)}(\theta,n)$ as

\begin{equation}
F_{(A,j)(\bar{A},k)}(\theta,n)=\begin{cases}
F_{(A,1)(\bar{A},1)}(2\pi i(k-j)-\theta,n) & \text{ if }k>j,\\
F_{(A,1)(\bar{A},1)}(2\pi i(j-k)+\theta,n) & \text{otherwise,}
\end{cases}\label{eq:FD2Full}
\end{equation}
and
\begin{equation}
F_{(A,j)(\bar{A},k)}(\theta,n)=F_{(\bar{A},j)(A,k)}(\theta,n)
\end{equation}
The FFs of the other field $\tilde{\mathcal{T}_n}$ denoted by $\tilde{F}$
can be simply obtained from those of $\TT_n$ \cite{Ola} through the relation
\begin{equation}
\tilde{F}_{(A,j)(\bar{A},k)}(\theta,n)=F_{(A,n-j)(\bar{A},n-k)}(\theta,n)\,.
\end{equation}

\subsection{Form Factors of the Disorder Operator $\mu_1$}
In this subsection we consider the form factor equations for the field $\mu_1$, the disorder operator associated to the flux $e^{i2\pi/3}$. This can be seen as a particular case of the composite twist field $\mathcal{T}^{\mu_1}_n$ when just a single replica is present ($n=1$). As mentioned before, we provide results for the disordered phase of the model, which are related to those in the ordered phase under the exchange $\mu_1 \leftrightarrow \sigma_1$ \cite{Potts_Delf}.

The disorder operators $\mu_{\pm 1}$ are $\mathbb{Z}_3$ invariant, so the form factors $F^{\mu_\pm}_{AA}(\theta)$ $F^{\mu_\pm}_{\bar{A}\bar{A}}(\theta)$ vanish by symmetry; from now on we focus one the particle-antiparticle form factors. The monodromy equation is
\be
F^{\mu_1}_{A\bar{A}}(\theta+2\pi i) = e^{\frac{2\pi i}{3}}F^{\mu_1}_{\bar{A}A}(-\theta),
\ee
since a mutual locality index $e^{i\alpha} = e^{ i2\pi/3}$ among $A$ and $\mu_1$ is present.
The unitarity equation is
\be
F^{\mu_1}_{\bar{A}A}(-\theta)S_{A\bar{A}}(\theta) = F^{\mu_1}_{A\bar{A}}(\theta)
\ee
A kinematical residue is present at $\theta = i\pi$ and it is related to the VEV $\la \mu_1\ra$ via
\be
\underset{\theta = i\pi}{\text{Res}} F^{\mu_1}_{A\bar{A}}(\theta) = i(1-e^{ \frac{2 \pi i}{3}}) \la \mu_1\ra,
\ee
while no dynamical poles are present\footnote{Note that this equation is different from the corresponding equation for BPTFs, where instead there are two separate kinematic residue equations. For local and semi-local fields though there is a single equation as explained for instance in \cite{Muss}.}. Similar equations hold if one exchanges $A\leftrightarrow \bar{A}$, keeping in mind that the only difference would be the mutual locality index $e^{-i2\pi/3}$ between $\bar{A}$ and $\mu_1$. A solution to these equations is given by\footnote{This solution is inspired from the solutions available for $U(1)$ fields in complex non-compact free boson, which is discussed in \cite{Ola-c1}}
\be
F^{\mu_1}_{A\bar{A}}(\theta) = -\la \mu_1 \ra \sin \frac{\pi}{3}\frac{e^{-\frac{\theta}{6}}}{\cosh \frac{\theta}{2}}  \frac{h(\theta,1)}{h(i \pi,1)},
\label{Fcompn1_PA}
\ee 
\be
F^{\mu_1}_{\bar{A}A}(\theta) = -\la \mu_1 \ra \sin \frac{\pi}{3}\frac{e^{\frac{\theta}{6}}}{\cosh \frac{\theta}{2}} \frac{h(\theta,1)}{h(i \pi,1)},
\label{Fcompn1_AP}
\ee
where the minimal form factor $h(\theta,1)$ is the $n=1$ case of the function analysed in Appendix \ref{minimal_FF}.

The structure of this solution can be easily justified: the minimal form factor solves the equations
\beq
h(\theta,1)=S(\theta)h(-\theta,1)=h(2\pi i-\theta,1)\,,
\eeq 
that is, the form factor equations in the absence of the semi-locality phase $e^{\frac{2\pi i}{3}}$. The factors  $r_\pm(\theta):=\frac{e^{\mp\frac{\theta}{6}}}{\cosh \frac{\theta}{2}} $ then account for this phase by satisfying
\beq
r_\pm(\theta)= r_\mp(-\theta)= e^{\pm \frac{2\pi i}{3}} r_\mp (2\pi i-\theta)\,,
\eeq  
and finally the denominator $\cosh\frac{\theta}{2}$ ensures the presence of a kinematic pole at $\theta=i\pi$ whilst the constant factors ensure the correct normalisation of the residue at the pole. 

For $\theta \rightarrow +\infty$ it is possible to show that
\be
h(\theta,1) \sim e^{\frac{|\theta|}{3}}, \quad |\theta| \rightarrow +\infty
\ee
thus, our solution has the following asymptotics
\be
\underset{\theta \rightarrow -\infty}{\lim} F^{\mu_1}_{A\bar{A}}(\theta) = \text{const.}
\label{muCluster}
\ee
This observation is in complete agreement with results of \cite{Potts_Delf,Zn_Parafermions} and the general form of the momentum space cluster property. In \cite{Potts_Delf}, it was shown that the FFs of the order operator satisfy the following cluster property in the ordered phase:
\be
\lim_{\theta \rightarrow \infty} \abs{F^{\sigma}_{1}(\theta)}=\frac{\left(F^\mu_K\right)^2}{\la \sigma\ra}\,,
\ee
where $K$ refers to a 1-kink excitation and $F^{\sigma}_1$ is a particular component of a 2-kink FF of the order operator.  Using \cite{Zn_Parafermions} and identifying the $\mathbb{Z}_3$ parafermion CFT with the 3-state scaling Potts model, one can easily show that the above limit in \eqref{muCluster} is proportional to the one-particle FF of the order operator in the disodered phase. Given the results of \cite{Potts_Delf,Zn_Parafermions} and making use of the Kramers-Wannier duality, one can infer that 
\be
\underset{\theta \rightarrow \infty}{\lim} \abs{F^{\mu_1}_{A\bar{A}}(\theta) }=\frac{\abs{F^{\sigma_1}_{\bar{A}}}^2}{\la \mu_1 \ra}\,.
\label{muClusterFinal}
\ee
The other FF $F^{\sigma_1}_{A}$ is vanishing due to the transformation properties of the field $\sigma_1$ and the particles $A,\bar{A}$ under the $\mathbb{Z}_3$ symmetry.

For the other disorder operator $\mu_{-1}$ similar considerations apply. A way to get easily the form factors of $\mu_{-1}$ is exploiting the charge-conjugation symmetry, meaning that $\mu_{-1}$ is the charge-conjugated field of $\mu_1$ and so its form factors can be obtained by interchanging $A\leftrightarrow \bar{A}$ in the previous formula. Similarly to the previous disorder field, the momentum space clustering can now be written as
\be
\underset{\theta \rightarrow \infty}{\lim} \abs{F^{\mu_{-1}}_{A\bar{A}}(\theta)} =\frac{\abs{F^{\sigma_{-1}}_{A}}^2}{\la \mu_{-1} \ra}\,.
\label{muClusterFinal2}
\ee
In general, solutions to the bootstrap equations are not unique, since a CDD ambiguity is present. 
However, in our case, thanks to clustering, we have enough constraints as to fix all parts of the form factor. It is still interesting though to employ the $\Delta$-theorem to test these solutions for consistency, using the ultraviolet features of the field $\mu_1$ under consideration. We will do this for the replica theory and the CTF $\mathcal{T}^{+}_n$, noting that the standard disorder operator is recovered for $n=1$.\\

\subsection{Form Factor Equations for Composite Branch Point Twist Fields\label{CBPTF}}

Relying on the exchange properties of the disorder CTFs \eqref{CTFSpatialExchange}
and also on earlier works \cite{Z2IsingShg,U(1)FreeFF, SGSRE} we can easily
write down the bootstrap equations for the novel composte twist fields.
Importantly, these equations include the non trivial phase $e^{i2\pi/3}$
in the monodromy properties corresponding the Aharonov-Bohm flux.
Unlike for continuous $U(1)$ symmetries discussed in \cite{U(1)FreeFF,SGSRE}, now we cannot divide the phase $2\pi/3$ by $n$ and distribute the flux uniformly on all the levels of the Riemann surface, as the division of phase is no longer compatible with the properties of the $\mathbb{Z}_3$ disorder fields. We can proceed in two possible ways. We either insert the same flux corresponding to the phase  $e^{i2\pi/3}$ on all the levels. This approach was applied in \cite{Z2IsingShg}, but in our current case, it is only legitimate for specific replica numbers $n=1,4,7,10,...$ where $e^{in2\pi/3}=e^{i2\pi/3}$ clearly holds. The other approach consist of introducing the flux in such a way, that the phase $e^{i2\pi/3}$ appears only when a particles moves from the $n$-th sheet to the $1$-th one. This choice introduces a slight asymmetry  between the replicas, but it is applicable for any positive number $n$. In this work, we follow this latter approach. Denoting the FFs of $\mathcal{T}_n^{\tau}$ by $F^{\tau}_{\underline{a}}(\underline{\theta},n)$, the
bootstrap equations can be formulated accordingly as

\begin{eqnarray}
 &  & F_{\underline{a}}^{\tau}(\underline{\theta},n)=S_{a_{i}a_{i+1}}(\theta_{i,i+1})F_{\ldots a_{i-1}a_{i+1}a_{i}a_{i+2}\ldots,n}^{\tau}(\ldots\theta_{i+1},\theta_{i},\ldots,n),\label{eq:U1FFAxiom1}\\
 &  & F_{\underline{a}}^{\tau}(\theta_{1}+2\pi i,\theta_{2},\ldots,\theta_{k},n)=F_{a_{2}a_{3}\ldots a_{k}\hat{a}_{1}}^{\tau}(\theta_{2},\ldots,\theta_{k},\theta_{1},n)\times \begin{cases}e^{i\kappa_{1}\tau2\pi/3} & \nu_1=n \\ 1 & \text{otherwise} \end{cases},\label{eq:U1FFAxiom2}\\
 &  & -i\underset{\theta_{0}'=\theta_{0}+i\pi}{{\rm Res}}F_{\bar{a}_{0}a_{0}\underline{a}}^{\tau}(\theta_{0}',\theta_{0},\underline{\theta},n)=F_{\underline{a}}^{\tau}(\underline{\theta},n),\label{eq:U1FFAxiom3}\\
 &  & -i\underset{\theta_{0}'=\theta_{0}+i\pi}{{\rm Res}}F_{\bar{a}_{0}\hat{a}_{0}\underline{a}}^{\tau}(\theta_{0}',\theta_{0},\underline{\theta},n)=-\prod_{l=1}^kS_{\hat{a}_{0}a_{l}}(\theta_{0l}) F_{\underline{a}}^{\tau}(\underline{\theta},n)\times \begin{cases}e^{i\kappa_{0}\tau2\pi/3} & \nu_0=n \\ 1 & \text{otherwise} \end{cases},\nonumber \\
 &  & -i\underset{\theta_{0}'=\theta_{0}+i\bar{u}_{\gamma \gamma}^{\bar{\gamma}}}{{\rm Res}}F^{\tau}_{(\gamma,\nu_{0})(\gamma,\nu'_{0})\underline{a}}(\theta_{0}',\theta_{0},\underline{\theta},n)=\delta_{\nu_{0},\nu'_{0}}\Gamma_{\gamma \gamma}^{\bar{\gamma}}F^{\tau}_{(\bar{\gamma},\nu_{0})\underline{a}}(\theta_{0},\underline{\theta},n),\label{eq:U1FFAxiom4}
\end{eqnarray}
recall that 
all notations are as for the BPTF discussed earlier.
The Lorentz spin $\Sigma=0$ as the CTF
is a spinless field too, moreover its $\mathbb{Z}_3$ charge is zero as well. The index $\kappa$ in the phase factors
corresponds to the $\mathbb{Z}_3$ charge of the corresponding particle, that
is,
\begin{equation}
\begin{split}\kappa_{i}= & \begin{cases}
1 & \gamma_{i}=A\\
-1 & \gamma_{i}=\bar{A}\,.
\end{cases}\end{split}
\end{equation}
The one-particle FFs are again vanishing, and the two-particle
FFs depend on the rapidity difference only. Like for  the BPTF, the novel
CTF is neutral in relation to  $\mathbb{Z}_3$ symmetry,
which implies the vanishing of any twist field FFs involving
a different number of particles and anti-particles mod 3. 

Under these considerations, Watson's equations \eqref{eq:U1FFAxiom1},
\eqref{eq:U1FFAxiom2} for non-vanishing two-particle form factors
and particles in the same copy can be summarised as 
\begin{eqnarray}
F_{a_{i}a_{j}}^{\tau}(\theta,n) & = & S_{a_{i}a_{j}}(\theta)F_{a_{j}a_{i}}^{\tau}(-\theta,n)=e^{i\kappa_{i}\tau2\pi/3}F_{a_{j}a_{i}}^{\tau}(2\pi in-\theta,n)\,.\label{fpm2U1}
\end{eqnarray}
The kinematic
residue equations \eqref{eq:U1FFAxiom3} are 
\begin{equation}
-i\underset{\theta=i\pi}{{\rm Res}}F_{a_{i}a_{i}}^{\tau}(\theta,n)=\langle\mathcal{T}^{\tau}_n\rangle\,\label{kinU1}
\end{equation}
where $\langle\mathcal{T}^{\tau}_n\rangle$ is the vacuum expectation
value of the CTF in the ground state of the replica theory. Finally we stress that the two-particle FFs with arbitrary replica
indices can be straightforwardly obtained form the above quantities
(corresponding to particles on the 1st replica only) as

\begin{equation}
F_{(A,j)(\bar{A},k)}^{\tau}(\theta,n)=\begin{cases}
F_{(\bar{A},1)(A,1)}^{\tau}(2\pi i(k-j)-\theta,n) & \text{ if }k>j,\\
F_{(A,1)(\bar{A},1)}^{\tau}(2\pi i(j-k)+\theta,n) & \text{ otherwise,}
\end{cases}
\label{eq:FCPTAAbarFull}
\end{equation}
and similarly for $A\leftrightarrow \bar{A}$, together with the important relation
\begin{equation}
F_{(A,1)(\bar{A},1)}^{\tau}(\theta,n)=F_{(\bar{A},1)(A,1)}^{-\tau}(\theta,n)\,.
\end{equation}
The FFs of
the other field $\tilde{\mathcal{T}}^{\tau}_n$ denoted by $\tilde{F}^{\tau}_{\underline{a}}(\underline{\theta},n)$
can be simply written as \cite{U(1)FreeFF}
\begin{equation}
\tilde{F}_{(\gamma,j)(\bar{\gamma},k)}^{\tau}(\theta,n)=F_{(\gamma,n-j)(\bar{\gamma},n-k)}^{-\tau}(\theta,n)\,,
\end{equation}
where $\gamma=A,\bar{A}$ and $\bar{\bar{A}}=A$.

\subsection{Two-Particle Form Factors of the Composite Twist Fields}

In this section we provide a solution to Watson's equations for the CTFs $\mathcal{T}_n^{\tau}$ and the two-particle form factors. We assume explicitly that $n$, the number of replicas, is greater than $1$ in what follows; nevertheless, the solutions we obtain are such that they converge to  Eqs. \eqref{Fcompn1_PA} and \eqref{Fcompn1_AP} when analytically continued to $n\rightarrow 1$.

Let us start with the form factor $F^{+}_{(A,1)(\bar{A},1)}(\theta,n)$, from which the other FFs can be easily obtained exploiting symmetries. Specifying the bootstrap equations to the two-particle form factors, one has
\be
F^{+}_{(A,1)(\bar{A},1)}(\theta+2i\pi n,n) = F^{+}_{(\bar{A},1)(A,1)}(-\theta,n)e^{2i\pi/3}, \quad F^{+}_{(\bar{A},1)(A,1)}(-\theta,n) S_{A\bar{A}}(\theta) = F^{+}_{(A,1)(\bar{A},1)}(\theta,n),
\ee
\be
\underset{\theta=i\pi}{\text{Res}} F^{+}_{(A,1)(\bar{A},1)}(\theta,n) = i\langle \mathcal{T}_n^{+}\rangle, \quad \underset{\theta=2in\pi-i\pi}{\text{Res}} F^{+}_{(A,1)(\bar{A},1)}(\theta,n) = -ie^{2i\pi/3}\langle \mathcal{T}_n^{+}\rangle.
\ee
A solution to these equations is given by 
\be
F^{+}_{(A,1)(\bar{A},1)}(\theta,n) = -i\langle \mathcal{T}_n^{+}\rangle\frac{e^{-\frac{\theta}{6n}}}{2n\sinh \frac{\theta+i\pi}{2n}\sinh \frac{\theta-i\pi}{2n}}\l e^{\frac{i\pi}{6n}}\sinh \frac{\theta-i\pi}{2n} - (\text{c.c.}) \r \frac{h(\theta,n)}{h(i\pi,n)},
\label{Fcompn_PA}
\ee
which, apart from the different pole structure, is very much reminiscent of the solution (\ref{Fcompn1_PA}). Here $h(\theta,n)$ is the minimal form factor discussed in Appendix \ref{minimal_FF}. 

Let us now comment briefly on the structure of the solution. The factor $e^{\frac{\theta}{3n}}h(\theta,n)$ ensures the right monodromy properties under $\theta \rightarrow \theta+2\pi i n$. The remaining factors, are  invariant under $\theta\rightarrow \theta+ 2\pi i n$ and  introduce two poles in the extended physical strip at $\theta = i\pi, 2in\pi - i \pi$, whose residues are $i\langle \mathcal{T}_n^{+}\rangle$ and $-ie^{2i\pi/3}\langle \mathcal{T}_n^{+}\rangle$. 

Similar conclusions hold for the form factor $F^{+}_{(\bar{A},1)(A,1)}(\theta,n)$ except for the kinematical residue value at $\theta=2in\pi -i \pi$ which is given by
\be
\underset{\theta=2in\pi -i \pi}{\text{Res}}F^{+}_{(\bar{A},1)(A,1)}(\theta,n) = -ie^{-2i\pi/3}\langle \mathcal{T}_n^{+}\rangle.
\ee
A solution of the bootstrap equation for $F^{+}_{(\bar{A},1)(A,1)}(\theta,n)$ is
\be
F^{+}_{(\bar{A},1)(A,1)}(\theta,n) = i\langle \mathcal{T}_n^{+}\rangle\frac{e^{\frac{\theta}{6n}}}{2n\sinh \frac{\theta+i\pi}{2n}\sinh \frac{\theta-i\pi}{2n}}\l e^{-\frac{i\pi}{6n}}\sinh \frac{\theta+i\pi}{2n} + (\text{c.c.}) \r \frac{h(\theta,n)}{h(i\pi,n)}.
\label{Fcompn_AP}
\ee
In the limit $n\rightarrow 1$ the poles at $\theta = i\pi,2i\pi-in\pi$ of the form factors $F^{+}_{(A,1)(\bar{A},1)}(\theta,n)$ and $F^{+}_{(\bar{A},1)(A,1)}(\theta,n)$, defined by Eqs. \eqref{Fcompn_PA} and \eqref{Fcompn_AP}, combine to produce a single pole with residue given by the sum of the two residues present for $n\neq 1$. As expected, the FFs  $F^{\mu_1}_{(A,1)(\bar{A},1)}(\theta)$, $F^{\mu_1}_{(\bar{A},1)(A,1)}(\theta)$ found above (see Eqs. \eqref{Fcompn1_PA},\eqref{Fcompn1_AP}) are recovered.

\begin{figure}[h!]
\centering
	\includegraphics[width=0.7\linewidth]{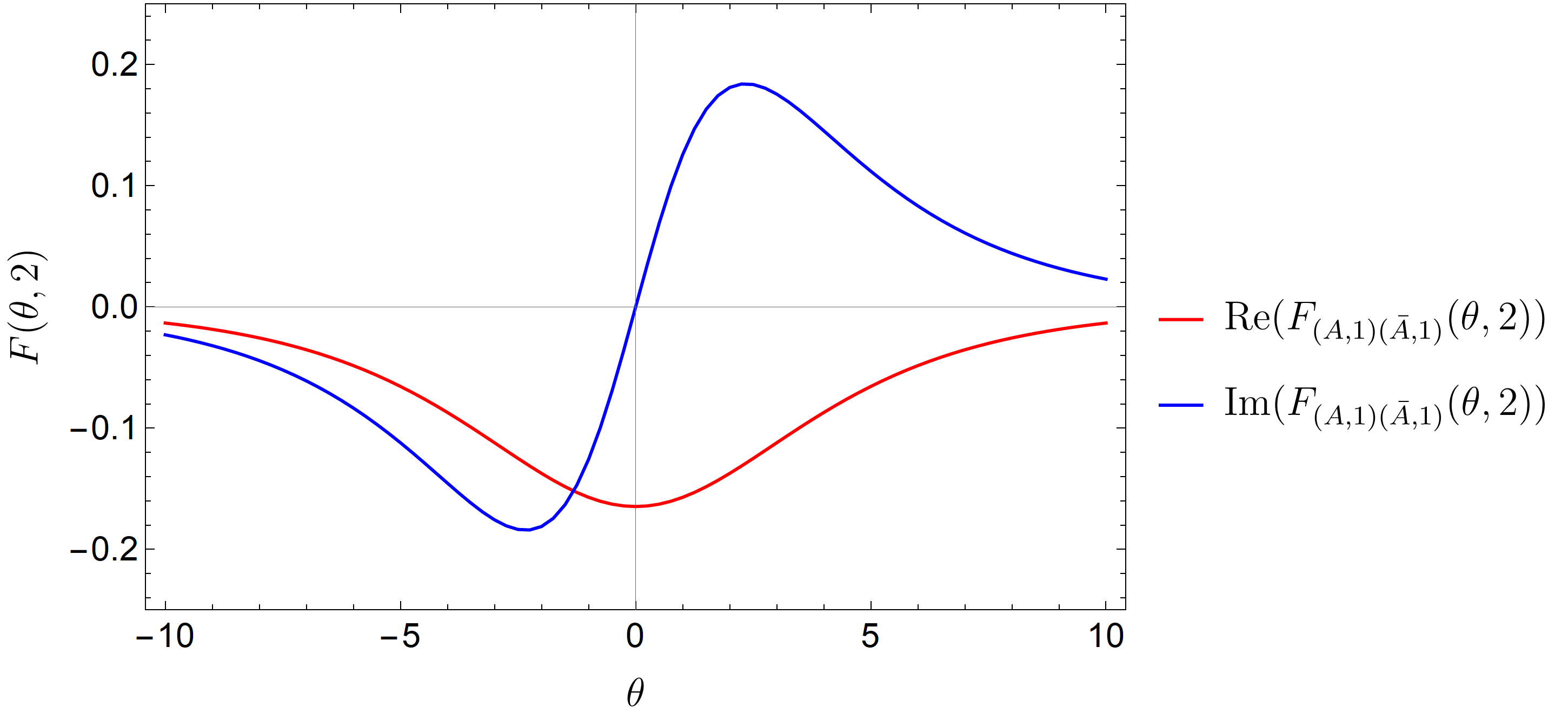}
	\caption{Two-particle form factors of the standard twist field $\mathcal{T}_n$ for $n=2$ replicas. 
	This figure shows the real and the imaginary parts of the $A\bar{A}$ form factor,  which equal those of the $\bar{A}A$ FF, 
	for real values of the rapidity difference $\theta$. They both go to zero when $\theta \rightarrow \pm \infty$.}
	\label{FF_standard}
\end{figure}	
\begin{figure}[h!]	
	\includegraphics[width=0.7\linewidth]{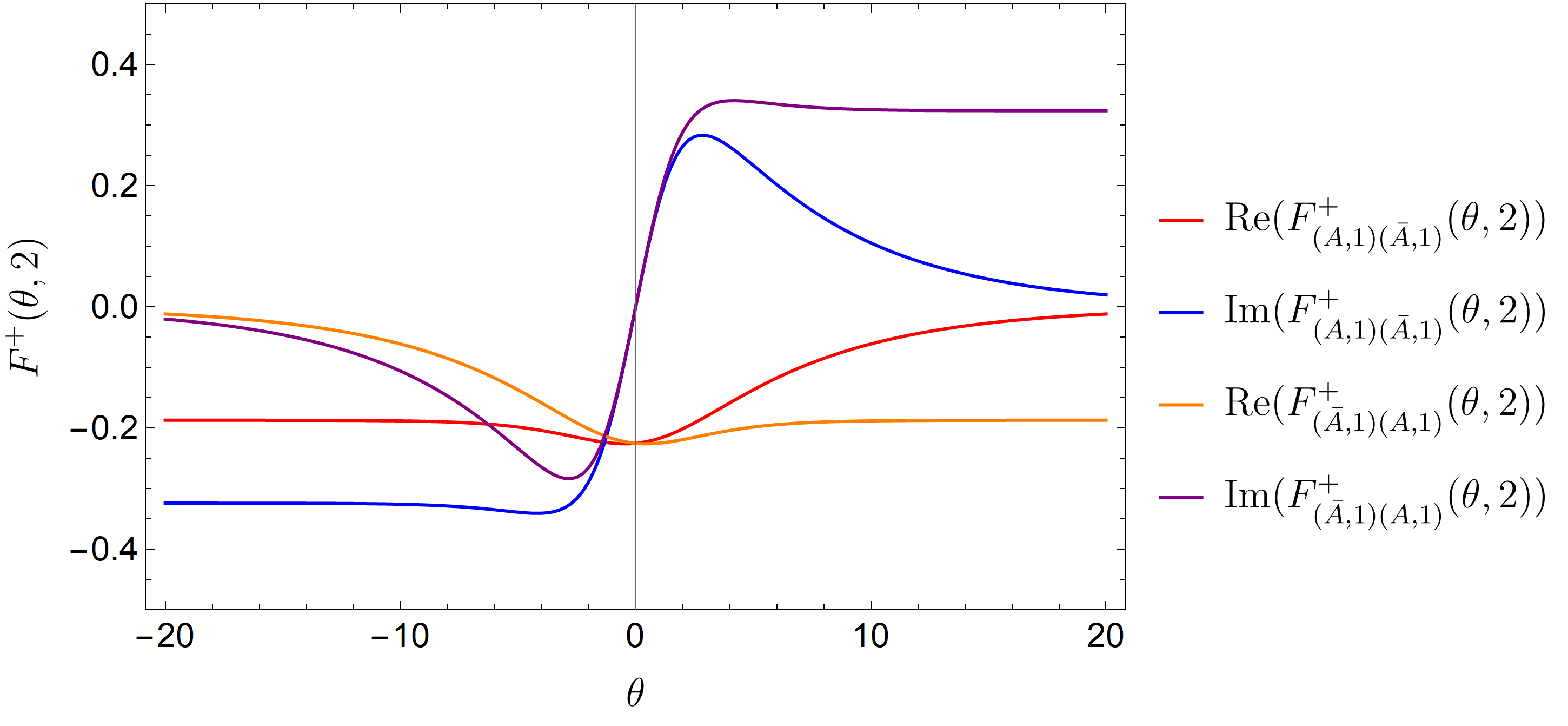}
	\caption{Two-particle form factors of the composite twist field $\mathcal{T}^{+}_n$ for $n=2$ replicas. This figure shows the $A\bar{A}$ and $\bar{A}A$ form factors, which are different, for real values of the rapidity difference $\theta$. They converge to a non-zero constant as $\theta \rightarrow \mp \infty$ respectively.}
	\label{FF_modified}
\end{figure}
In Figs. \ref{FF_standard} and \ref{FF_modified} we show the behaviour of the two-particle form factor for the BPTF and CTF respectively as a function of $\theta$ for $n=2$; our plots are normalized by the VEV of the twist fields. While for the BPTFs the correspondent FF is vanishing in the limit $\theta \rightarrow \pm \infty$, the FF of CTFs takes a non-zero asymptotic value as expected from clustering. 

\subsection{Checks via $\Delta$-Sum Rule}

In this subsection we check the compatibility of our CTF FFs with the predictions of the $\Delta$-theorem. We recall that the bootstrap equations generally admit many solutions, so some additional information is generally needed to identify a particular solution with a particular field. For this purpose, the $\Delta$-theorem \cite{delta_theorem} gives a non-trivial constraint which typically rules out unwanted solutions. It states that for any spinless operator $\mathcal{O}$ which is mutually local w.r.t. the trace of the stress-energy tensor $\Theta$ we have
\be
\Delta_{\mathcal{O}} = -\frac{1}{4\pi \langle \mathcal{O}\rangle}\int d^2x \langle \Theta(x)\mathcal{O}(0)\rangle_{\text{conn}.},
\ee
with $\Delta_{\mathcal{O}}$ the conformal dimension of the field $\mathcal{O}$ and $\langle \Theta(x)\mathcal{O}(0)\rangle_{\text{conn}.}$  the connected correlation function between $\Theta$ and $\mathcal{O}$. In principle on can expand  $\langle \Theta(x)\mathcal{O}(0)\rangle_{\text{conn}.}$ as an infinite sum over the quasiparticle basis, and the knowledge of the infinitely many form factors of $\mathcal{O}$ and $\Theta$ should be needed to reconstruct the correlation function. This is an extremely difficult task which has been performed explicitly only for free theories. There is however strong evidence from numerous works that the sum is quickly converging and the first terms thus provide the main contribution to the result (see e.g. \cite{Ola} for the BPTF of the sinh-Gordon model). We report here just the contribution coming from the two-particle FF, since the vacuum contribution is subtracted in the connected correlation function is considered and one-particle form factors of both fields vanish. We further assume, as is the case for the 3-state Potts model, that all the particles have the same mass $m$ so that we can write
\be
\Delta_{\mathcal{O}} \simeq  -\frac{1}{32\pi^2 \la \mathcal{O}\ra m^2} \int^{\infty}_{-\infty} d\theta \frac{1}{\cosh^2 \theta/2} \sum_{a,a'} F^{\Theta}_{a;a'}(\theta)\bar{F}^\mathcal{O}_{a;a'}(\theta).
\label{h_dtheorem}
\ee
The latter sum runs over the set of particles, labeled by internal indices $a,a'$. Eq. \eqref{h_dtheorem} straighforwardly generalizes to the replicated model, so that additional replica indices appear. One can thus apply the $\Delta$-theorem to the composite twist field $\mathcal{T}_{n}^{\tau}$, obtaining
\be
\Delta_{n}^\tau \simeq  -\frac{n}{32\pi^2 \la \mathcal{T}_{n}^{\tau}\ra m^2} \int^{\infty}_{-\infty} d\theta \frac{1}{\cosh^2 \theta/2} \sum_{\gamma,\gamma'} F^{\Theta}_{(\gamma,1)(\gamma',1)}(\theta)\bar{F}^{\tau}_{(\gamma,1)(\gamma',1)}(\theta,n),
\ee
where we used the fact that the two-particle FFs of $\Theta$ vanish for different replica indices as all copies are independent \cite{Ola}.
Let us now focus on the 3-state Potts model. First, we just have to consider two species of particles $A,\bar{A}$ with the same mass $m$, related by charge conjugation. The only non-vanishing two-particle FFs of the stress-energy tensor are
\be
F^{\Theta}_{(A,1)(\bar{A},1)}(\theta)  = F^{\Theta}_{(\bar{A},1)(A,1)}(\theta) = 2\pi m^2 \frac{h(\theta,1) }{h( i\pi ,1)}
\ee
and they are normalised such that $F^{\Theta}_{(A,1)(\bar{A},1)}(i \pi) = 2\pi m^2$. Thus, the $\Delta$-theorem gives
\be
\Delta_n^\tau \simeq  -\frac{n}{32\pi^2 m^2\la \mathcal{T}^{\tau}_{n}\ra} \int^{\infty}_{-\infty} d\theta \frac{1}{\cosh^2 \theta/2} (F^{\Theta}_{(A,1)(\bar{A},1)}(\theta)\bar{F}^{\tau}_{(A,1)(\bar{A},1)}(\theta,n) + F^{\Theta}_{(\bar{A},1)(A,1)}(\theta)\bar{F}^{\tau}_{(\bar{A},1)(A,1)}(\theta,n) ).
\label{delta_potts}
\ee
This result has to be compared with the CFT prediction, which reads as follows \cite{bcd-15,gs-18}
\be
\Delta_{{n}}^\tau = \Delta_n + \frac{\Delta_{\tau}}{n} \quad \mathrm{with}\quad \Delta_n=\frac{c}{24}\l n-\frac{1}{n}\r \,,
\label{dimT}
\ee
where $\Delta_n$ and $\Delta_\tau$ are the conformal dimensions of the BPTF \cite{k-87,dixon} and of $\mu_\tau$, respectively.

\begin{figure}[h!]
\centering
	\includegraphics[width=0.9\linewidth]{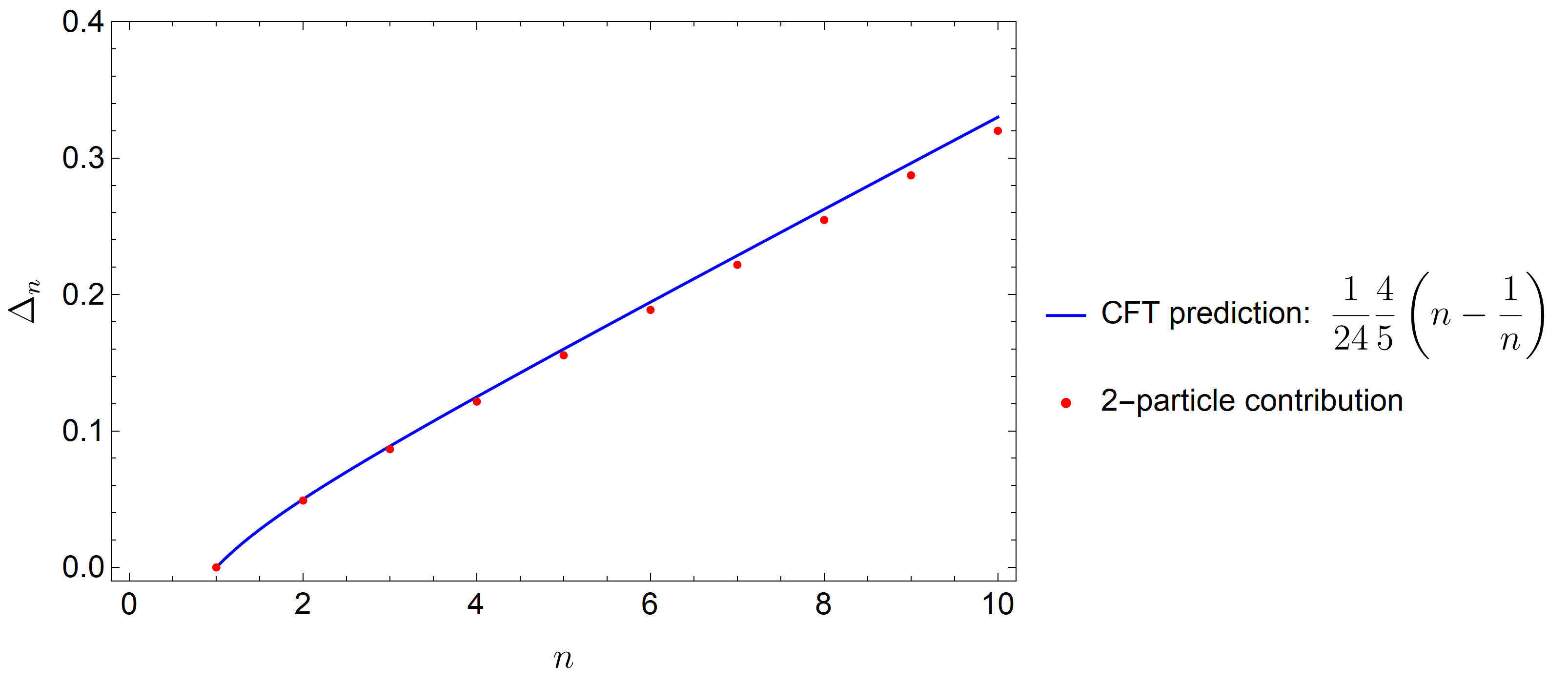}
	\caption{Conformal dimension of the BPTF $\mathcal{T}_n$ as obtained from the $\Delta$-theorem (red dots) compared to the exact CFT formula (blue dots).}
	\label{h_standard}
	\end{figure}

\begin{figure}[h!]
	\includegraphics[width=0.9\linewidth]{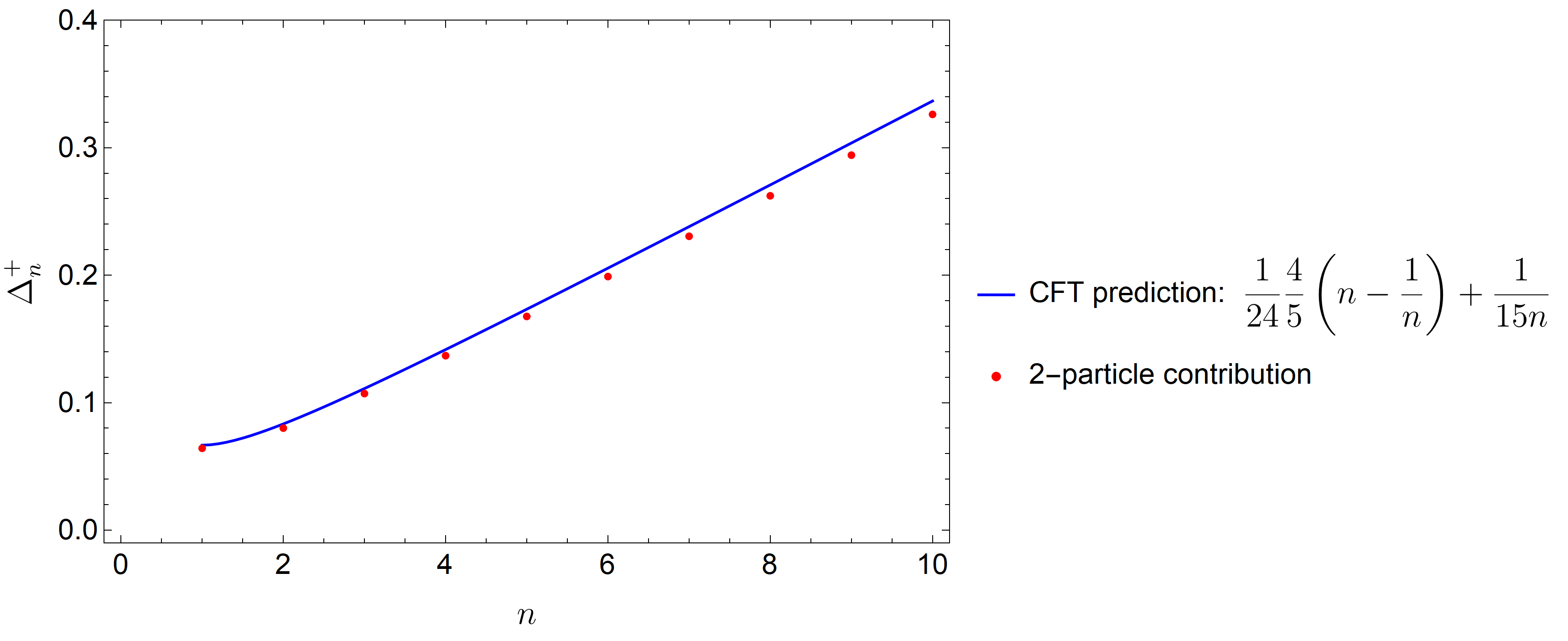}
	\caption{Conformal dimension of the CTF $\mathcal{T}^{\tau}_n$ as obtained from the $\Delta$-theorem (red dots) compared to the exact CFT formula (blue dots).}
	\label{h_mod}
\end{figure}

For the 3-state Potts model the value of the central charge is $c=4/5$ and the dimensions 
\be
\Delta_{\tau} = \begin{cases} \frac{1}{15} \quad \tau=\pm1,\\ 
0 \quad \tau=0,\end{cases}
\ee
where we remind ourselves that the CTFs can be associated with the specific values $\alpha=\tau 2\pi/3$ of the Aharonov-Bohm flux.
Figs. \ref{h_standard} and \ref{h_mod} show this comparison for standard and composite twist fields respectively, where the integral appearing in Eq. \eqref{delta_potts} has been performed numerically. The data are very close to the CFT value, but discrepancies are expected to be present since we are neglecting higher particle form factors.


\section{$\mathbb{Z}_3$ Charged Moments}

\begin{figure}[h!]
\floatbox[{\capbeside\thisfloatsetup{capbesideposition={right,top},capbesidewidth=4cm}}]{figure}[\FBwidth]
{ \caption{$2f_{A\bar{A}}(0,n,\tau)$ is shown as a function of $n$ both for the standard ($\tau=0$) and CTFs ($\tau=\pm 1$). As expected, its asymptotic value in the limit $n\rightarrow \infty$ does not depend on $\tau$.}
\label{f_0_n}} 
{
	 \includegraphics[width=12cm]{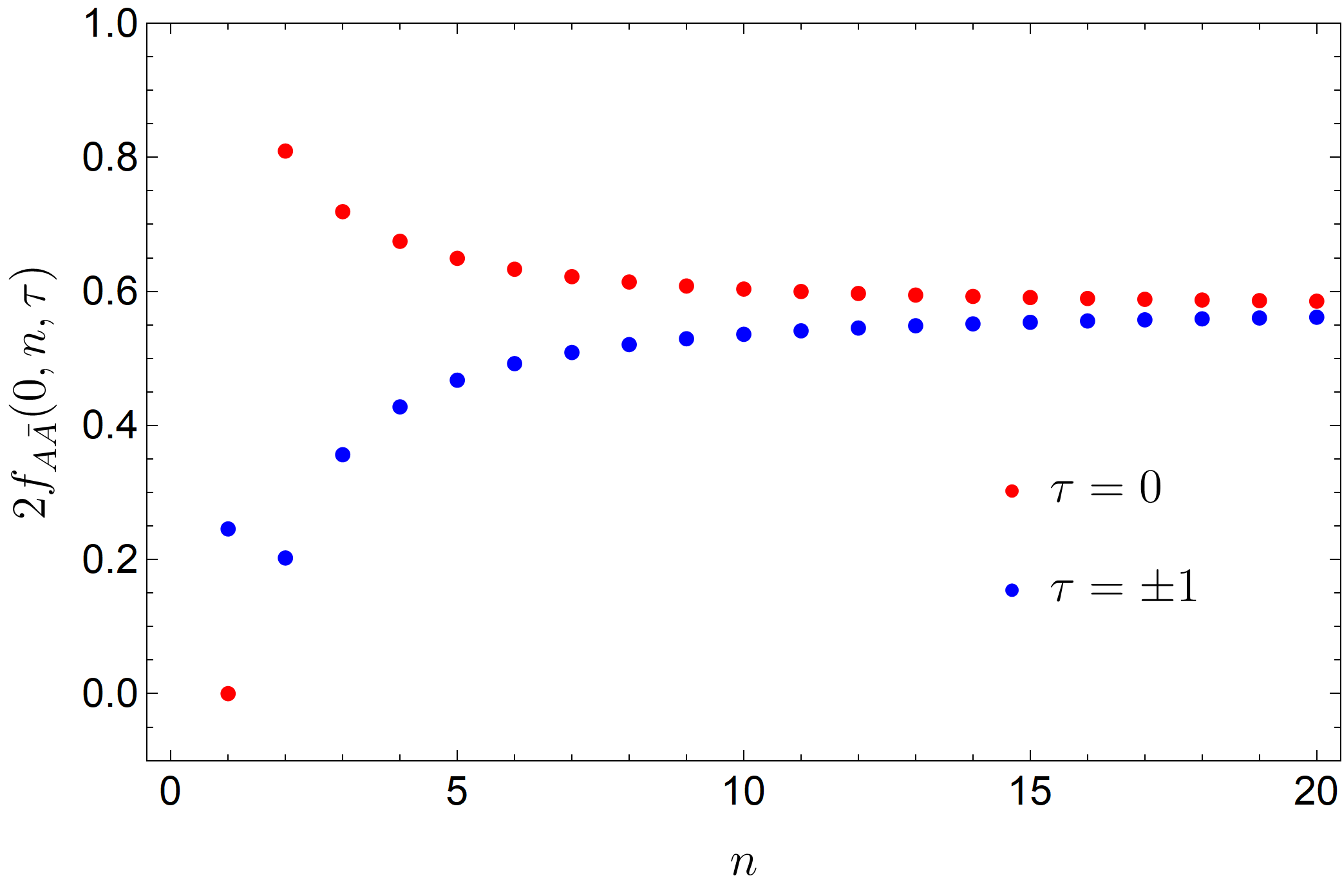}}
	\end{figure}

In this section we analyze the long-distance behaviour of the correlation function of CTFs using our predictions related to their two-particle FFs. The latter quantity is strictly related to the one interval symmetry resolved R\'enyi entropy, which is the main focus of our work. For now, we keep the values of $\alpha$ generic and only at the end specify what happens for the $\mathbb{Z}_3$ resolution. To do so, first of all we have to identify the scaling limit of the charged moments (defined by Eq. \eqref{cmoments}) which is
\be
Z_n(\alpha) \equiv \trace\l \rho_A^n e^{i\alpha Q}\r = \epsilon^{4\Delta_{n}^\tau} \langle \TT^\tau_{n}(0)\tilde{\TT}^\tau_{n}(\ell)\rangle \quad \mathrm{with} \quad \alpha=\frac{2\pi \tau}{3}\,,
\label{Zn_alpha}
\ee
for the subsystem $[0,\ell]$. The constant $\epsilon$ plays the role of UV regulator, and we suppose that $\ell,m^{-1}\gg \epsilon$ so that the scaling limit is reached. Secondly, we expand the correlation function $\langle \mathcal{T}^\tau_{n}(0)\tilde{\mathcal{T}^\tau_{n}}(\ell)\rangle$ in the quasiparticle basis keeping  only the two-particle contribution
\begin{equation}
\begin{split}\langle\mathcal{T}_{n}^{\tau}(0)\tilde{\mathcal{T}}_{n}^{\tau}(\ell)\rangle\simeq & \langle\mathcal{T}_{n}^{\tau}\rangle^{2}+\sum_{j,k=1}^{n}\int_{-\infty}^{\infty}\frac{\mathrm{d}\theta_{1}\mathrm{d}\theta_{2}}{(2\pi)^{2}2!}|F^\tau_{(A,j)(\bar{A},k)}(\theta_1-\theta_2,n)|^{2}e^{-\ell m\left(\cosh\theta_{1}+\cosh\theta_{2}\right)}\\
 & \qquad+\sum_{j,k=1}^{n}\int_{-\infty}^{\infty}\frac{\mathrm{d}\theta_{1}\mathrm{d}\theta_{2}}{(2\pi)^{2}2!}|F^\tau_{(\bar{A},j)(A,k)}(\theta_1-\theta_2,n)|^{2}e^{-\ell m\left(\cosh\theta_{1}+\cosh\theta_{2}\right)}\\
= & \langle\mathcal{T}_{n}^{\tau}\rangle^{2}\left(1+\frac{n}{4\pi^{2}}\sum_{\gamma,\gamma'=A,\bar{A}}\int_{-\infty}^{\infty}\mathrm{d}\theta f_{\gamma \gamma'}(\theta,n,\tau)K_{0}\left(2m_\gamma \ell\cosh\left(\theta/2\right)\right)\right)\,.
\end{split}
\label{corr_CTF}
\end{equation}
Above,  $K_0(z)$ is the modified Bessel function and the functions $f_{\gamma {\gamma'}}(\theta,n,\tau)$ are implicitly defined as
\be
\begin{split}
f_{A\bar{A}}(\theta,n,\tau)\langle\mathcal{T}_{n}^{\tau}\rangle^{2} =\sum_{j=1}^{n}|F^\tau_{(A,1)(\bar{A},j)}(\theta,n)|^{2}
= \sum_{j=0}^{n-1}|F^\tau_{(A,1)(\bar{A},1)}(\theta-2i\pi j,n)|^{2}\,,\end{split}
\ee
and similarly for $f_{\bar{A}A}(\theta,n,\tau)$. By  symmetry in the current theory $f_{AA}(\theta,n,\tau)=f_{\bar{A}\bar{A}}(\theta,n,\tau)=0$. Note that for the 3-state Potts model $m_A=m_{\bar{A}}=m$.
Then, expanding the integral in Eq. \eqref{corr_CTF} at leading order in the limit $m\ell \rightarrow \infty$ we get
\be
\langle\mathcal{T}_{n}^{\tau}(0)\tilde{\mathcal{T}}_{n}^{\tau}(\ell)\rangle\simeq \langle\mathcal{T}_{n}^{\tau}\rangle^{2} \l 1+\frac{n}{2\pi} f_{A\bar{A}}(0,n,\tau)\frac{e^{-2m\ell}}{m\ell}\r.
\ee
where we used the fact that $f_{A\bar{A}}(0,n,\tau)=f_{\bar{A}A}(0,n,\tau)$. 
For $n=1$ and $\tau=\alpha=0$ the twist field becomes the identity operator and $f_{A\bar{A}}(\theta,1,0)=f_{\bar{A}A}(\theta,1,0)=0$. However, for $\tau\neq 0$ we have that $f_{A\bar{A}}(\theta,1,\pm 1)\neq 0$, including for $\theta=0$. Thus it is this contribution which dictates the leading behaviour of the correlation function for the Potts model. In Fig. \ref{f_0_n}
 we plot $f_{A\bar{A}}(0,n,\tau)+f_{\bar{A}A}(0,n,\tau)=2f_{A\bar{A}}(0,n, \tau)$ as a function of $n$, for $n$ integer, for both $\tau=0$ and $\tau=\pm 1$ (which are identical). From the figure it is evident that $f_{A\bar{A}}(0,n,\tau)$ converges to a $\tau$-independent constant when $n\rightarrow \infty$. We were able to compute analytically this constant (see Appendix \ref{Analf} for the details). The result is
  \be
2f_{A\bar{A}}(0,\infty,\tau) =\frac{16}{\pi^2}\frac{\Gamma^2\l 5/6\r \Gamma^2\l 2/3\r}{\Gamma^2\l 4/3\r \Gamma^2\l 1/6\r}\l    _{5}F_{4}\left[ \begin{array}{c}
     -1/2, -1/2, 5/6, 5/6, 1 ; 1 \\
    1/2, 1/2, 3/2, 3/2 \\
   \end{array}\right] -1/2\r \simeq 0.575153\,,
\ee
and the derivation is very similar to the sinh-Gordon analysis presented in \cite{Ola}.

\subsection{Analytic Continuation $n\rightarrow 1$}

The analytic continuation of $f_{A\bar{A}}(\theta,n,\tau)$ for non-integer values of $n$, needed for the computation of the symmetry resolved von Neumann entropy, is a nontrivial task. In this subsection we discuss its universal features for $n\rightarrow 1$ for a generic IQFT. We specialize our results to the 3-state Potts model at the end of the subsection. A careful analysis of the Ising and sinh-Gordon model has been performed in \cite{Ola, cd-09} in absence of flux ($\tau=\alpha=0$), while the dependence on $\alpha$ for non-interacting $U(1)$ free theories was discussed in \cite{U(1)FreeFF}. The main subtlety is occurrence of singularities in the limit $n\rightarrow 1$, coming from the collision of kinematical poles leading to non-uniform convergence and a derivative $\frac{d}{dn}f_{A\bar{A}}(\theta,n,\tau)|_{n=1}$  which becomes singular. It has been shown that for any 1+1D QFT the following limit holds in absence of flux
\be
\underset{n\rightarrow 1}{\lim} \frac{d}{dn}f_{\gamma\bar{\gamma}}(\theta,n,0) = \frac{\pi^2}{2}\delta(\theta).
\ee
An important feature of the previous expression is the independence on the precise form of the $S$-matrix and the presence of the $\delta$-singularity, not shared for other values of $n$. When the flux is present one expects $\frac{d}{dn}f_{\gamma\bar{\gamma}}(\theta,n,\tau)|_{n=1}$ to consist of a $\delta$-like term, singular for $\theta=0$, and a smooth $\theta$-dependent term. For example, for free fermionic/bosonic $U(1)$ theories, carefully studied in \cite{U(1)FreeFF}, where two species of particles ($\gamma,\bar{\gamma}$) related by charge conjugation are present it has been found that
\be
\underset{n\rightarrow 1}{\lim} \frac{d}{dn}f_{\gamma\bar{\gamma}}
(\theta,n,\tau) = \frac{\pi^2}{2}\cos \alpha\, \delta(\theta)+ (\text{smooth $\theta$-dependent part})\,,
\ee
where $\alpha$ is the $U(1)$ charge. 
The smooth $\theta$-dependent term is different for bosons and fermions while the singular part is the same for both theories. We now show that the same general mechanism occurs also for interacting theories with discrete symmetries. Let us consider again the function $f_{\gamma\gamma'}(\theta,n,\tau)$ in an interacting IQFT, which is defined as before 
\be
\begin{split}
f_{\gamma\gamma'}(\theta,n,\tau)\langle\mathcal{T}_{n}^{\tau}\rangle^{2} =\sum_{j=1}^{n}|F^\tau_{(\gamma,1)(\gamma',j)}(\theta,n)|^{2} 
=  \sum_{j=0}^{n-1}|F^\tau_{(\gamma,1)(\gamma',1)}(\theta-2i\pi j,n)|^{2},
\end{split}
\ee
we claim that
\be
\underset{n\rightarrow 1}{\lim} \frac{d}{dn}f_{\gamma{\gamma'}}(\theta,n,\tau) = \frac{\pi^2}{2} \cos \alpha_\gamma \, \delta(\theta) \delta_{\gamma'\bar{\gamma}}+ (\text{smooth $\theta$-dependent part}).
\label{univ_alpha}
\ee
Here $e^{i\alpha_\gamma}$ is the Aharonov-Bohm phase of species $\gamma$ introduced by the CTF $\mathcal{T}_n^\tau$. The proof goes in the same way as for $U(1)$ free theories in \cite{U(1)FreeFF}. For instance, one writes $f_{\gamma\gamma'}(\theta,n,\tau)$ as
\be
f_{\gamma\gamma'}(0,n,\tau)\langle\mathcal{T}_{n}^{\tau}\rangle^{2} = \sum_{j=0}^{n-1}|F^\tau_{(\gamma,1)(\gamma',1)}(\theta-2i\pi j,n)|^{2} = \sum_{j=0}^{n-1} s_{\gamma\gamma'}(\theta,j,\tau),
\ee
which implicitly defines $s_{\gamma\gamma'}(\theta,j,\tau)$. Then, the sum over $j$ is performed by considering the following contour integral
\be
\frac{1}{2\pi i}\oint_{\mathcal{C}} dz \pi \cot \l \pi z \r s_{\gamma\gamma'}(\theta,z,\tau),
\label{circ_int}
\ee
where the contour is a rectangle with vertices $(-\epsilon-iL,n-\epsilon-iL,n-\epsilon+iL,-\epsilon+iL)$. The integral can be computed by residue theorem; it has poles at $z=0,1,\dots,n-1$, present for any pair $(\gamma,\gamma')$, and at $z=\frac{1}{2}\pm \frac{\theta}{2\pi i}$ and $z= n-\frac{1}{2}\pm \frac{\theta}{2\pi i}$, coming from the kinematic poles of the particle-antiparticle form factors (that is, from terms with $\gamma'=\bar{\gamma}$). Using the explicit values of the residues, we end up with
\be
\begin{split} 
\frac{1}{2\pi i }\oint_{\mathcal{C}} dz \, \pi \cot \l \pi z \r s_{\gamma\gamma'}(\theta,z,\tau) = f_{\gamma\gamma'}(\theta,n,\tau)\langle\mathcal{T}_{n}^{\tau}\rangle^{2} +
\\
 \frac{\tanh^2 \frac{\theta}{2}}{\la \mathcal{T}_n^\tau\ra^2} \text{Im}\l F^{\tau}_{(\gamma,1) (\bar{\gamma},1)}(i\pi-2\theta,n)-e^{-i\alpha_\gamma}F^{\tau}_{(\gamma,1)(\bar{\gamma},1)}(i\pi(2n-1)-2\theta,n)\r \delta_{\gamma'\bar{\gamma}}.
\end{split}
\ee
The integral is in general hard to compute, but if we just focus on the singular limit $n\rightarrow 1$ for which the kinematic poles collide, then we get
\begin{eqnarray}
f_{\gamma\gamma'}(\theta,n,\tau) &=& -\frac{\tanh^2 \frac{\theta}{2}}{\la \mathcal{T}_n^\tau\ra^2}\sum_\gamma \text{Im}\l F^{\tau}_{(\gamma,1)( \bar{\gamma},1)}(i\pi-2\theta,n)-e^{-i\alpha_\gamma}F^{\tau}_{(\gamma,1)(\bar{\gamma},1)}(i \pi(2n-1)-2\theta,n)\r \delta_{\gamma'\bar{\gamma}} \nonumber\\
 && + (\text{reg. terms}).
\end{eqnarray}
Taking the derivative in $n$ of the previous expression and then the limit $n\rightarrow 1$, the expression in Eq. \eqref{univ_alpha} is finally obtained. If one specializes this prediction to the 3-state Potts model, where just $2$ species of particles $A$ and $\bar{A}$ are present, then for the nontrivial values of the flux $\alpha = \pm 2\pi/3$, that is $\tau=\pm 1$
\be
\underset{n\rightarrow 1}{\lim} \frac{d}{dn} f_{A\bar{A}}(\theta,n,\pm 1) = \frac{\pi^2}{2} \cos \frac{2\pi}{3}\delta(\theta) + (\text{smooth $\theta$-dependent part}).
\ee
A remark on Eq. \eqref{univ_alpha} is probably needed. As we stated before, it encodes the contribution of the two-particle form factors only and its singular behaviour does not really depend on the details of the S-matrix. Assuming no one-particle contributions one would conclude, after inserting Eq. \eqref{univ_alpha} in Eq. \eqref{corr_CTF}
\be
\lim_{n\rightarrow 1}\frac{d}{dn} \langle \mathcal{T}^{\tau}_{n}(0)\tilde{\mathcal{T}}_n^{\tau}(\ell)\rangle  \simeq \frac{1}{8}\sum_\gamma \cos \alpha_\gamma \, K_0(2m_\gamma \ell) + (\ell- \text{indep. terms}) + \dots,
\ee
with $m_\gamma$ being the mass of particle $\gamma$. However, in generic QFTs there can also be one-particle contributions, which would be leading for large distances. In the presence of nonzero flux typically survive also in the limit $n\rightarrow 1$ and so they really dictate the dominant behaviour of the symmetry resolved von Neumann entropy. An explicit example for this mechanism has been carefully discussed in the sine-Gordon model in Ref. \cite{SGSRE}.

\subsection{$\mathbb{Z}_3$ Symmetry Resolved Entropies}

In this subsection we analyze the scaling limit of the $\mathbb{Z}_3$ symmetry resolved R\'enyi entropy in the 3-state Potts model, using the correlation function among twist fields. We mention that similar analysis has been already performed \cite{Zn} for the same model in the critical regime, which correponds to the limit of vanishing mass $m\rightarrow 0$. Here we instead are interested in the long distance behaviour, for instance the limit of $m\ell \rightarrow \infty$ with $m$ kept fixed. To perform the calculation we express the symmetry resolved partition function $\mathcal{Z}_n(q)$, defined by Eq. \eqref{Zn_q} as
\be
\mathcal{Z}_n(q) \equiv  \tr\l \rho_A^n \Pi_q \r = \frac{1}{3}\sum^{1}_{j=-1}e^{-iq2\pi j/3}Z_n \l \frac{2\pi j}{3}\r, \quad q=0,\pm 1.
\ee
Inserting Eq. \eqref{Zn_alpha}, which relates composite twist fields and charged moments, one obtains
\be
\mathcal{Z}_n(q) = \frac{1}{3}\epsilon^{4\Delta_{n}^\tau}\l \la \mathcal{T}_{n}(0)\tilde{\mathcal{T}_{n}}(\ell)\ra +2\cos \l \frac{2\pi}{3}q\r\epsilon^{\frac{4}{n}\Delta_{+}}\la \mathcal{T}^{+}_{n}(0)\tilde{\mathcal{T}}_{n}^{+}(\ell)\ra\r,
\ee
since by charge conjugation symmetry of the state $Z_n(2\pi/3) = Z_n(-2\pi/3)$. If one keeps only the leading term in the definition of $\mathcal{Z}_n(q)$ in the limit $\epsilon \rightarrow 0$, the term coming from $Z_n(0)$, then we can approximate
\be
\mathcal{Z}_n(q) \simeq \frac{1}{3}Z_n(0).
\ee
At this order of analysis, the symmetry resolved entropies becomes
\be
S_n(q) \simeq S_n - \log 3,
\label{equip}
\ee
and the equipartition of the charge is recovered, meaning that there is no explicit dependence on the charge for the different charge sectors. Beyond this approximation, a dependence on $q$ is found. Now we investigate the first corrections to Eq. \eqref{equip} coming from the lowest powers of the regulator $\epsilon$. For $n>1$ we obtain the following
\be
S_n(q) = S_n -\log 3 +\frac{1}{1-n} 2\cos \frac{2\pi q}{3} \epsilon^{\frac{4}{n}\Delta_{+}} \frac{\la \mathcal{T}^{+}_{n}(0)\tilde{\mathcal{T}}_{n}^{+}(\ell)\ra}{\la \mathcal{T}_{n}(0)\tilde{\mathcal{T}_{n}}(\ell)\ra} + o( \epsilon^{\frac{4}{n}\Delta_{+}}).
\ee
The case $n\rightarrow 1$ has to be analyzed carefully, since logarithmic corrections appear. Let us thus expand $S_1(q)$, keeping just the correction of order $O(\epsilon^{4\Delta_{+}})$ and $O(\epsilon^{4\Delta_{+}}\log \epsilon)$
\begin{eqnarray}
S_1(q) &=& -\partial_n \log \mathcal{Z}_n(q)|_{n=1} + \log \mathcal{Z}_1(q) \nonumber \\
&=& S_1 -\log 3 -2\cos\frac{2\pi q}{3} \epsilon^{4\Delta_{+}} \la \mu_1 (0)\mu_{-1}(\ell)\ra  \\
&& \times \l -4\Delta_{+}\log \epsilon +\partial_n \log \la \mathcal{T}^{+}_{n}(0)\tilde{\mathcal{T}}_n^{+}(\ell)\ra|_{n=1} - \partial_n\log \la \mathcal{T}_{n}(0)\tilde{\mathcal{T}_{n}}(\ell)\ra|_{n=1} -1 \r + o(\epsilon^{4\Delta_{+}}). \nonumber
\end{eqnarray}
Although in the previous expression enters the normalization constants of the fields, which have not been fixed, they contribute just as $\ell$-independent constant. Moreover since we are mostly interested in the large $m\ell$ limit, we can just keep
\be
\partial_n \log \la \mathcal{T}^{+}_{n}(0)\tilde{\mathcal{T}}_n^{+}(\ell)\ra|_{n=1} - \partial_n\log \la \mathcal{T}_{n}(0)\tilde{\mathcal{T}_{n}}(\ell)\ra|_{n=1}\,,
\label{cont}
\ee
as the only $\ell$-dependent term; the reason is that the next-to-leading order large $\ell$ correction to $\la \mu_{1}(0) \mu_{-1}(\ell)\ra$ is of order
\be
 \la \mu_{1}(0)\mu_{-1}(\ell)\ra \simeq \la \mu_{1}\ra^2 + O\l \frac{e^{-2m\ell}}{m\ell}\r.
\ee
whereas the contributions from (\ref{cont}) are proportional to $K_0(2m\ell) \approx \frac{e^{-2m\ell}}{\sqrt{m\ell}}$.
Therefore, at leading and next-to-leading order for large $\ell$ we can write
\begin{multline}
S_1(q) = S_1 -\log 3 -2\cos \frac{2\pi q}{3} \epsilon^{4\Delta_{+}} |\la \mu_{1} \ra|^2\\
\times \l -4\Delta_{+}\log \epsilon +\frac{1}{4}\l \cos \frac{2\pi}{3}-1\r K_0(2m\ell) +(\ell-\text{indep. terms})\r + o\l \epsilon^{4\Delta_{+}},\frac{e^{-2m\ell}}{\sqrt{m\ell}}\r.
\end{multline}
In this last result we explicitly use our universal prediction, valid for large $m\ell$ in the two-particle approximation
\be
\frac{d}{dn} \left[ \frac{\la \mathcal{T}^{+}_{n}(0)\tilde{\mathcal{T}}_n^{+}(\ell)\ra}{\la\mathcal{T}^{+}_{n}\ra^2}\right]_{n=1} \simeq \frac{1}{4}\cos \frac{2\pi}{3}K_0(2m\ell).
\ee

\section{Conclusions}

In this work we apply the form factor bootstrap approach to computing the two-particle form factors of  composite branch point twist fields in the 3-state Potts model, an integrable quantum field theory with diagonal scattering. Although the 3-state Potts model in the disordered phase is the main focus of our work, we also provide some results for the $q$-state Potts model in the ordered phase with $q\leq 4$, focusing on the (more standard) branch point twist field.

Our main analytical result is an expression for the leading and next-to-leading large-distance behaviour of the symmetry resolved R\'enyi and von Neumann entropies. The same technique has been applied in Ref. \cite{Z2IsingShg} to the Ising model and the sinh-Gordon model, generalized to $U(1)$ free theories in Ref. \cite{U(1)FreeFF} and to sine-Gordon in Ref. \cite{SGSRE}. The 3-state Potts model  is arguably the simplest interacting model displaying a non-trivial $\mathbb{Z}_3$ internal symmetry and therefore of particular interest in the context of symmetry resolved entanglement. Despite the presence of interaction, many technical details are still shared with simpler integrable theories, such as diagonal scattering or the vanishing of one-particle form factors for composite twist fields. 

Even if the main goal of this work was the symmetry resolved entropy,  the standard R\'enyi entropy had also not been previously studied for the 3-state (or more generally $q$-state) Potts model, and this the reason why in Appendix A we also provided explicitly the form factors of the standard branch point twist field and some standard results for the entanglement entropy. In this context, perhaps the main result is the observation that the next-to-leading order correction to the von Neumann entropy of a large region $\ell$ is, as for other models, given universally in terms of a modified Bessel function $K_0(2m\ell)$, where $m$ is the mass of the $q-1$ lightest particles (kinks, in the ordered phase) in the theory. The coefficient of this Bessel function is indeed proportional to $q-1$.

All our form factor solutions have been tested for consistency, employing the standard $\Delta$-theorem (or $\Delta$ sum rule) and the limit $n\rightarrow \infty$, as well as the analytical continuation to $n\rightarrow 1$, have been considered. In this manner, we were able to also show that the next-to-leading term of the symmetry resolved von Neumann entropy does not really depend on the details of the $S$-matrix. Indeed, as for the total entropy, it is proportional to a modified Bessel function $K_0(2m\ell)$ with a coefficient that is dictated by the internal symmetry and particle content. The result is easily generalizable to other theories with vanishing one-particle form factors and discrete internal symmetries. 

Finally, it is worth mentioning that the symmetry resolved R\'enyi entropy of the critical 3-state Potts has been studied in Ref. \cite{Zn}, employing conformal field theory techniques. Our approach is complementary, since it gives access to the properties of the model away from criticality. Moreover, despite focussing only on the one-interval geometry in the large-size limit, in principle at least, the knowledge of all the higher-particle form factors (which remains an open problem) would give access to all possible geometries of the subsystem. This is indeed a possible subject for further investigations. One could also apply these techniques to investigate the role of the internal symmetry in other entanglement measurements, e.g. the negativity or the relative entropy. As already mentioned, our analysis can be slightly adapted to deal with other integrable models displaying internal symmetries as the integrable perturbations of the parafermionic $\mathbb{Z}_N$ theories \cite{fz-91}.

\subsection*{Acknowledgments}
LC, DXH, and PC acknowledge support from ERC under Consolidator grant number 771536 (NEMO). LC is grateful to Gesualdo Delfino for fruitful discussions.

\begin{appendix}

\section{Form Factors and Entanglement Entropy in the Ordered Phase}
\label{FFBPTF}
In \cite{Potts_Delf} the two-particle form factors of the stress-energy tensor and order and disorder fields in the ordered phase were computed. Although in order to consider BPTFs we must first work on a replica version of the model, in essence the solution procedure follows very closely the work of Delfino and Cardy for the stress-energy tensor as the twist field is also a ``neutral'' operator with respect to the internal symmetry of the model. Thus, the only non-vanishing for factors would be those corresponding to neutral states, as defined in the Introduction. Let $K_{(\alpha, i) (\beta,i)}(\theta)$ represent a kink traveling between vacuum $\alpha$ and vacuum $\beta$ in copy $i$ of the replica theory. The two-particle form factor associated to two kinks living on copy 1 of the theory is the object
\beq 
F(\theta_1-\theta_2,n)=\langle 0_\alpha|\mathcal{T}_n(0)| K_{(\alpha,1) (\beta,1)}(\theta_1) K_{(\beta,1) (\alpha,1)}(\theta_2)\rangle. 
\eeq 
Due to the permutation symmetry under vacua exchange in the original model, the form factors will not depend on $\alpha, \beta$ as long as $\alpha \neq \beta$.
The form factor above will share many properties with its counterpart for the stress-energy tensor, but also has some differences: first, it is now a function of $n$ and, second, contrary to the stress-energy tensor, it must have kinematic residue poles which are specified by the twist field form factor equations (\ref{eq:FFAxiom1})-(\ref{eq:FFAxiom4}). As usual, a minimal solution for $g(\theta,n)$ can be obtained by solving the unitarity and crossing conditions:
\beq 
g(\theta,n)=\Lambda(\theta) g(-\theta,n)=g(-\theta + 2\pi i n, n).
\eeq 
where 
\beq 
\Lambda(\theta)=(q-2)S_2(\theta)+S_3(\theta)=-\exp \left(\int_0^\infty \frac{dt}{t} f(t) \sinh\frac{t\theta}{i\pi} \right),
\eeq 
with 
\beq 
f(t)=\frac{2}{\sinh \frac{t}{2\lambda}}\left(\sinh t\left(\frac{1}{2\lambda}-1\right)+ \frac{\sinh\frac{t}{3} \cosh\frac{t}{2}\left(\frac{1}{3}-\frac{1}{\lambda}\right)}{\cosh\frac{t}{2}}\right).
\eeq 
A minimal solution is given by
\beq 
g(\theta,n)=-i \sinh\frac{\theta}{2n} \exp\left(\int_0^\infty \frac{dt}{t \sinh(nt)} f(t) \sin^2\left(\frac{it}{2}\left(n+\frac{i\theta}{\pi}\right) \right)\right)\,,
\label{gfmin}
\eeq 
It is interesting to notice that for $\lambda=1$ 
\beq 
f(t)=2 \left(-1+\frac{\sinh\frac{2t}{3}}{\sinh t}\right)\,,
\eeq
and then (\ref{gfmin}) coincides with the solution (\ref{Fmin}) for the minimal form factor in the disordered phase at $q=3$. This is because 
\beq
-i \sinh\frac{\theta}{2n}= \exp\left(2\int_0^\infty \frac{dt}{t \sinh(nt)} \sin^2\left(\frac{it}{2}\left(n+\frac{i\theta}{\pi}\right)\right)\right)\,,
\label{imp}
\eeq
which means that the $-1$ term in $f(t)$ effectively cancels the factor $-i \sinh\frac{\theta}{2n}$ in (\ref{gfmin}). 

\subsection{Form Factors and $\Delta$-Sum Rule for $\lambda < 1$ ($q < 3$)}
For the BPTF it makes sense to consider values of $q$ that are not necessarily integer. In this section we compute FF solutions for  $q< 3$ in the ordered phase.  In this case there are $q< 3$ kinks with identical scattering matrices and form factors. There are no bound states. The minimal form factor is given above and the full two-particle form factor is then determined by the kinematic requirement of having poles at $\theta=i\pi$ and $\theta=i\pi(2n-1)$. This gives
\beq 
F(\theta,n)=\frac{\langle \mathcal{T}_n \rangle \sin\frac{\pi}{n}}{2n \sin\frac{i\pi-\theta}{2n}\sin\frac{i\pi+\theta}{2n}} \frac{g(\theta,n)}{g(i\pi,n)}.\label{ff}
\eeq 
Form factors involving other copies can be obtained from this by the usual identities presented earlier in the paper. This FF solution can be checked against the $\Delta$-sum rule as we show below. Recall first of all that the conformal dimension of the BPTF was given  in (\ref{dimT}) and that,  in the Potts model the UV central charge $c$ is a function of $q$ (or $\lambda$) given by
\beq 
c=1-\frac{6}{t(t+1)} \quad \text{with} \quad \frac{2\lambda}{3}=\frac{t-1}{t+1}.
\eeq 
Note that for $\lambda<\frac{1}{2}$ the central charges are negative. 

\begin{figure}[h!]
\begin{center}
\includegraphics[width=6.9cm]{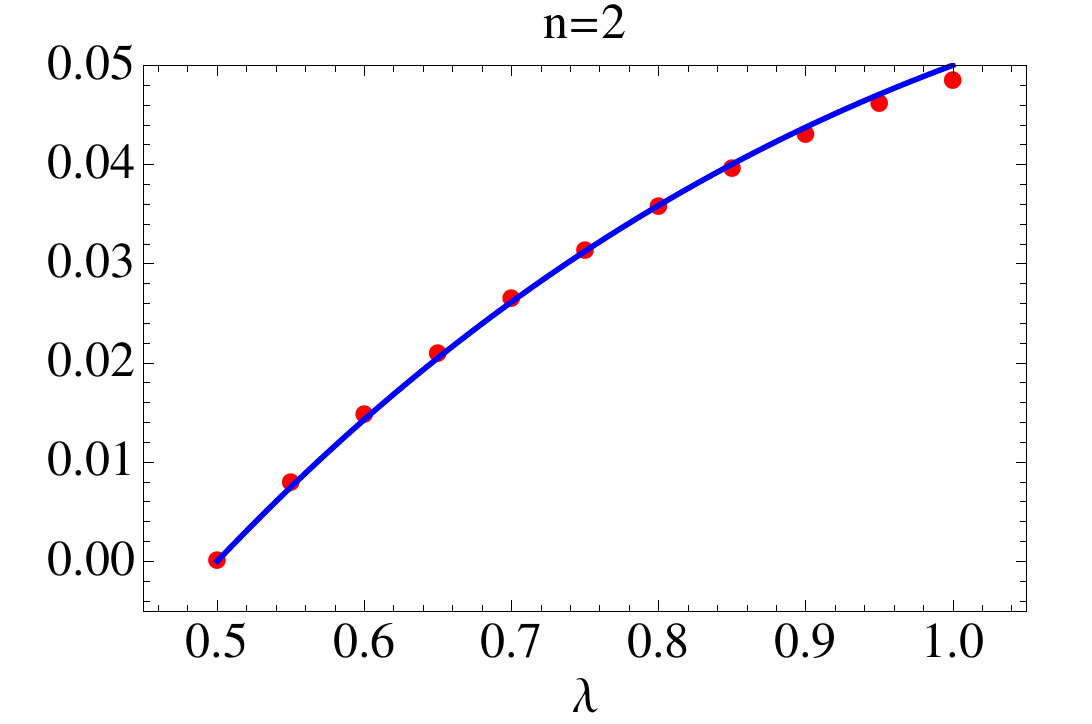}
\includegraphics[width=6.9cm]{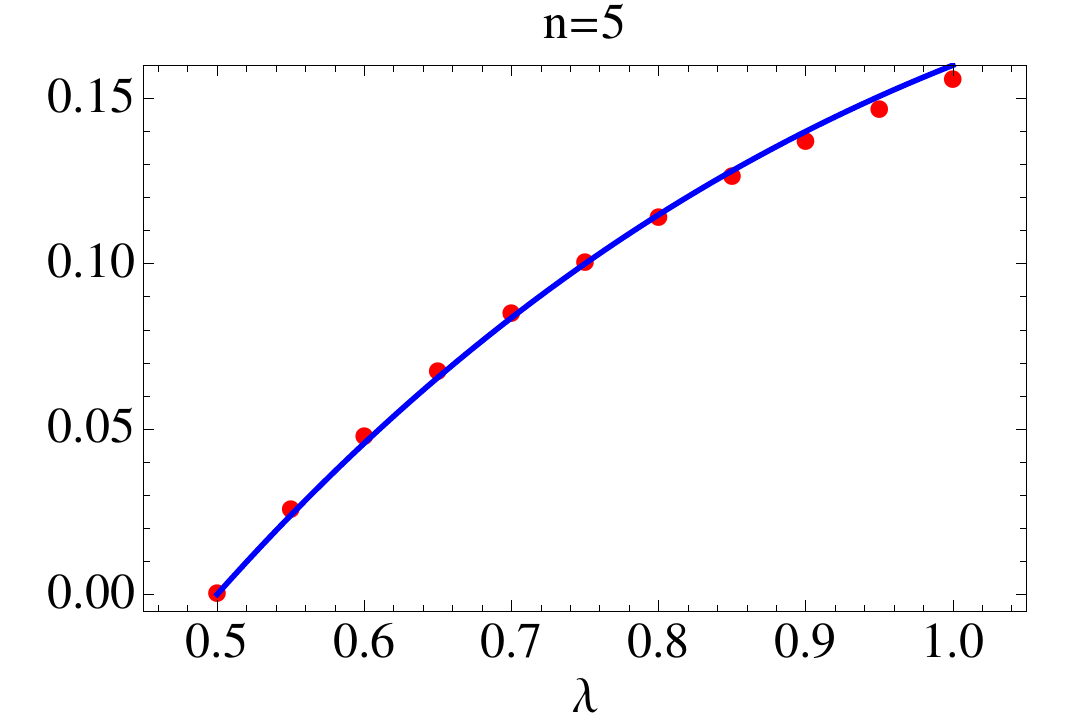} 
\includegraphics[width=6.9cm]{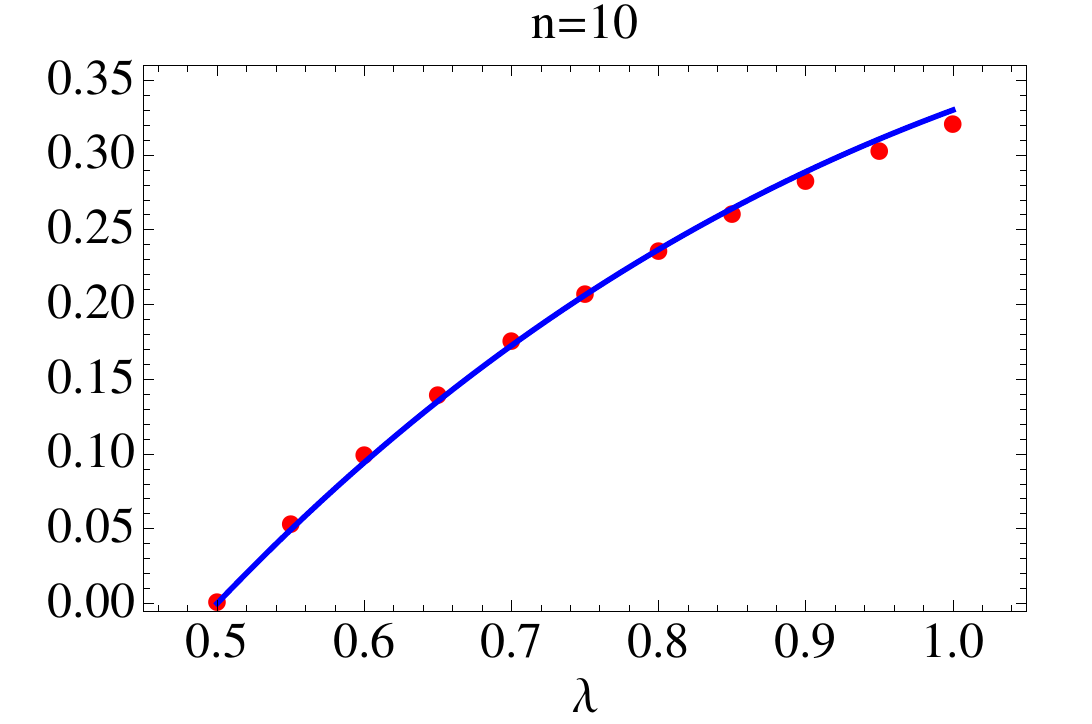} 
\includegraphics[width=6.9cm]{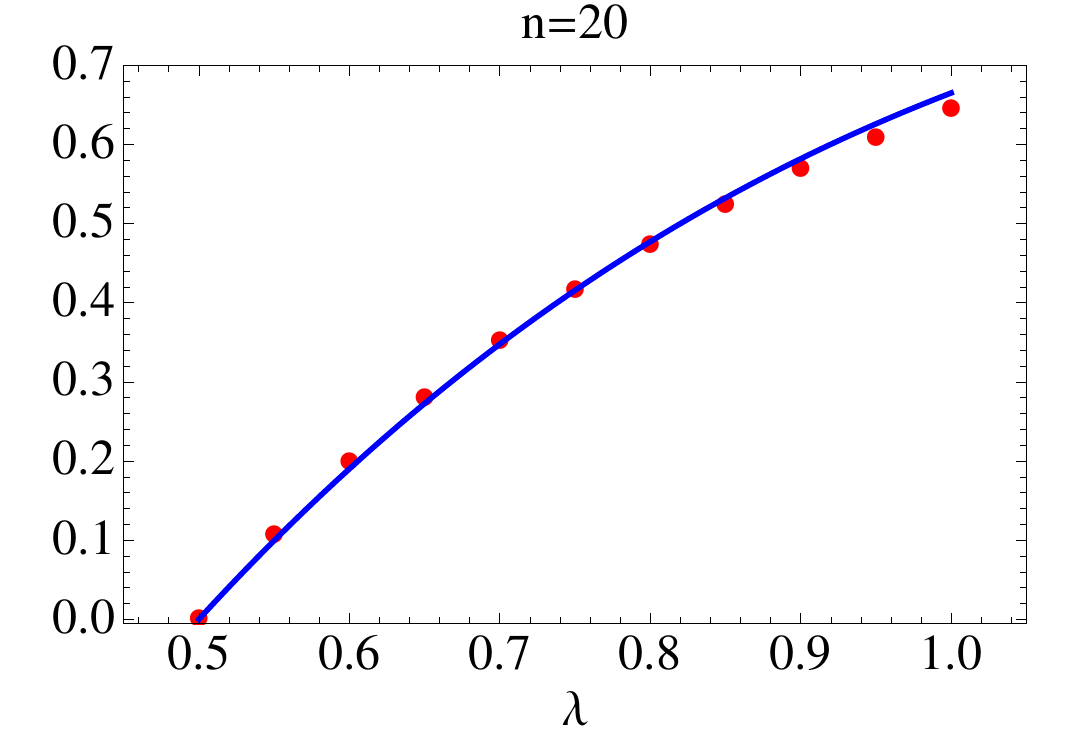} 
\caption{$\Delta$-sum rule in the two-particle approximation (red dots) compared to the exact formula (solid blue line) for $n=2, 5, 10, 20$ and $\frac{1}{2} \leq \lambda \leq 1$.}
\label{f1}
\end{center}
\end{figure}

The two-particle form factor of $\Theta$ was computed in \cite{Potts_Delf} and in our notation it is simply given by
\beq 
G(\theta)=\langle 0_\alpha|\Theta(0)| K_{(\alpha,1) (\beta,1)}(\theta_1) K_{(\beta,1) (\alpha,1)}(\theta_2)\rangle=2\pi m^2 \frac{g(\theta,1)}{g(i \pi, 1)}.
\eeq 
We may expand the two-point function above and truncate the expansion at the level of the two-particle form factors. This is generally quite simple but in this case we must take care to consider the fact that we are now working with $n$ copies of the Potts model (this will result on a factor $n$) and that we must also sum over intermediate states (this will result on a factor $q-1$), after simplification we obtain 
\beqa 
\Delta_n \approx  -\frac{n (q-1)}{8 m^2(2\pi)^2 \langle\mathcal{T}_n\rangle}  \int_{-\infty}^\infty dx  \frac{F(x,n) G(x)^{*}}{\cosh^2\frac{x}{2}}.
\eeqa 
A numerical evaluation of this integral is presented in Fig.~\ref{f1}. Despite the approximation, agreement with the exact formula (\ref{dimT}) is extremely good for all values of $\lambda$, including $\lambda=1$.
\subsection{Form Factors for $1 < \lambda\leq \frac{3}{2}$}
As noted earlier, for $\lambda > 1$ the minimal form factor $g(\theta,n)$ must be analytically continued and in addition account must be taken of the presence of the additional bound state pole at $\theta_B$ (\ref{thetaB}). The solution involves a new minimal form factor
\beq 
\tilde{g}(\theta,n)=-i a(\theta,n) \sinh\frac{\theta}{2n} \exp\left(\int_0^\infty \frac{dt}{t \sinh(nt)} \tilde{f}(t) \sin^2\left(\frac{it}{2}\left(n+\frac{i\theta}{\pi}\right) \right)\right),
\eeq 
where $a(\theta, n)$ encodes the bound state pole
\beq 
a(\theta,n)=\frac{\cosh^2\frac{\theta_B}{2n}}{\cosh\frac{\theta_B}{n}-\cosh\frac{\theta}{n}},
\eeq 
at $\theta=\theta_B$ and 
\beq 
\tilde{f}(t)=\frac{2}{\sinh \frac{t}{2\lambda}}\left(\sinh t\left(\frac{3}{2\lambda}-1\right)+ \frac{\sinh\frac{t}{3} \cosh\frac{t}{2}\left(\frac{1}{3}-\frac{1}{\lambda}\right)}{\cosh\frac{t}{2}}\right)\,.
\eeq 
It is interesting to analyse what happens to these formulae at $\lambda=1$.
For $\lambda=1$ we have that $\theta_B=0$ and so $a(\theta,n)=(1-\cosh\frac{\theta}{n})^{-1}$ gives a pole at $\theta=0$. Overall 
\beq
a(\theta,n) \sinh\frac{\theta}{2n} \mapsto \frac{1}{\sinh\frac{\theta}{2n}}\,.
\eeq
This pole is however cancelled by the minimal form factor. This is due once more to the relation (\ref{imp}) and the fact that this precise integral results from the contribution of the term $\sinh t\left(\frac{3}{2\lambda}-1\right)$ in $\tilde{f}(t)$ for $\lambda=1$. Thus, $g(\theta,n)=\tilde{g}(\theta,n)$ for $\lambda=1$ and they both coincide with (\ref{Fmin}), the minimal form factor in the disordered phase which is discussed in Appendix \ref{minimal_FF}. 

\subsubsection{Entanglement Entropy}
As we have seen, the focus of this paper is the computation of the symmetry resolved entanglement entropy. Here we present a very brief summary of the main properties of the total entropy for the Potts model with $q\leq 3$ as would follow from the form factor above. Similar to the study of sinh-Gordon and Ising presented in \cite{Ola}, the next-to-leading order correction to saturation of both the R\'enyi entropy associated to a parameter $n\in \mathbb{Z}^+$ and the von Neumann entropy are given by the two-particle form factor contribution to the the expansion of the two-point function of BPTFs. For the von Neumann entropy, the result can be written straight away as it follows from the general results of \cite{Ola,next}. We have that
\beq 
S(\ell)=\frac{c}{3}\log(m^{-1}\epsilon)+U-\frac{q-1}{8}K_0(2m\ell)+\cdots 
\eeq 
where $m$ is the mass of the $q-1$ kinks present in the model (assuming they all have the same mass), $\ell$ is the size of the interval, $U$ is a constant related to the expectation value of the twist field and $\epsilon$ is a non-universal short distance cut-off which is chosen so that at the critical point the entanglement entropy scales as $S(r)=\frac{c}{3}\log\frac{\ell}{\epsilon}$ and there are no further constant corrections. An interesting feature, which is particular to the Potts model is that the number of kinks $q$ emerges as a prefactor of the Bessel function. Thus the entanglement entropy encodes basic information about the particle content of the theory.

The R\'enyi entropy $S_n(\ell)$ will also acquire exponentially decaying corrections on the size of the sub-system but they will take a less universal form and depend on the details of the two-particle form factor. By definition, we have that 
\beq 
S_n(\ell)=\frac{1}{1-n}\log(A_n \epsilon^{-4\Delta_n} \langle \mathcal{T}_n(0) \tilde{\mathcal{T}}_n(\ell) \rangle). 
\eeq 
$A_n$ is a non-universal constant with the property $A_1=1$, $\Delta_n$ is the conformal dimension of the twist fields given earlier and the two-point function now can be expanded in terms of the form factors we have computed before. Employing the relationships between form factors of particles living of different replicas we obtain
\beqa 
\log {\langle \mathcal{T}_n(0) \tilde{\mathcal{T}}_n(\ell) \rangle} &\approx & \frac{n (q-1)}{ 2(2\pi)^2} \int_{-\infty}^\infty  \int_{-\infty}^\infty d\theta_1 d\theta_2 e^{-\ell m(\cosh \theta_1 + \cosh \theta_2)} \sum_{j=0}^{n-1} |F^{11}(\theta_2-\theta_1+2\pi i j,n)|^2 \nonumber \\
&=& \frac{n (q-1)}{(2\pi)^2} 
\int_{-\infty}^\infty   d x  \sum_{j=0}^{n-1} |F^{11}(-x+2\pi i j,n)|^2 K_0(2m\ell\cosh\frac{x}{2}):=s(\ell,n).
\eeqa 
Then, the R\'enyi entropy may be written as
\beq 
S_n(\ell)=\frac{n+1}{6n}\log(m^{-1}\epsilon)+U_n+\frac{s(\ell,n)}{1-n}+\cdots 
\eeq 
where we have parameterized $\langle \mathcal{T}_n \rangle^2 =m^{4\Delta_n} v_n$ so that $U_n=\frac{\log(v_n A_n)}{1-n}$. The function $s(\ell,n)$ can be easily evaluated numerically for integer $n$ or otherwise analytically continued to real values of $n$ along the lines of \cite{Ola}.

\section{Minimal Form Factor in the Disordered Phase}\label{minimal_FF}
In this appendix we analyize the behavior of the minimal form factor $h(\theta,n)$ in the replica 3-state Potts model.
As in the previous appendix and following closely \cite{Ola}, we start with an integral representation of the $S$-matrix
\be
S_{A\bar{A}}(\theta) = -\frac{\sinh \l \frac{\theta}{2}+i\frac{\pi}{6} \r}{\sinh \l \frac{\theta}{2}-i\frac{\pi}{6} \r} =\exp \l \int_0^\infty \frac{dt}{t} \sinh \frac{t\theta}{i\pi} \frac{2 \sinh \l \frac{2}{3}t\r}{\sinh t}\r\,.
\ee
Then, the only solution (up to a multiplication constant) which does not have zeros or poles in the region $\text{Im} \theta \in (0,2\pi n)$ and satisfies the bootstrap axioms 
\be
h(\theta+2\pi i n, n) = h(-\theta, n), \qquad h(-\theta,n) S_{A\bar{A}}(\theta) = h(\theta,n)\,,
\ee
is given by
\beq
h(\theta,n) = \exp \left[ \int_0^\infty \frac{dt}{t\sinh (nt)} \frac{2 \sinh\frac{2t}{3}}{\sinh t} \sin^2 \l \frac{it}{2}\l n +\frac{i\theta}{\pi}  \r \r \right].
\label{Fmin}
\eeq
It is easy to prove that for real $\theta$ the integral is convergent, since the for $t\rightarrow \infty$ it scales as $t^{-1}{e^{-t/3}}$ 
and no singularity is present in the limit $t\rightarrow \infty$. However, for $\theta = -i|\theta|$  the integrand goes as  $t^{-1} e^{-t/3+t|\theta|/\pi}$ for large $t$, that is, it is no longer convergent for $\text{Im}(\theta) \leq \pi/3$. 

For real $\theta$ in the limit $\theta \rightarrow +\infty$ a more careful but standard estimate of the integral gives
\be
 \int_0^\infty \frac{dt}{t\sinh (nt)} \frac{2 \sinh \l \frac{2}{3}t\r}{\sinh t} \sin^2 \l \frac{it}{2}\l n +\frac{i\theta}{\pi}  \r \r \simeq \frac{4}{3n}\int_0^{\infty} \frac{dt}{t^2}\sin^2 \frac{t \theta}{\pi}  = \frac{\theta}{3n}. 
\ee
The approximate equality holds at order $O(\theta)$, and so the asymptotic growth of the minimal form factor is
\be
h(\theta,n) \sim e^{\frac{\theta}{3n}}, \qquad \theta \rightarrow +\infty.
\ee
Finally, we would like to present the mixed product representation of $h(\theta,n)$ that we have used in our numerical work. Expanding\footnote{We can also choose to expand the factor $1/\sinh (nt)$, leading to an equivalent representation.} $1/\sinh t$ as
\be
\frac{1}{\sinh t} = 2e^{-t}\sum^{N-1}_{n=0}e^{-2nt} + \frac{e^{-2Nt}}{\sinh t}
\ee
and using the integral representation of the $\Gamma$-function
\be
\Gamma(z) = \exp \left( \int_0^\infty \frac{dt}{t} \l\frac{e^{-tz}-e^{-t}}{1-e^{-t}}+(z-1)e^{-t}\r \right),
\ee
the integral \eqref{Fmin} gives, after a lengthy but straightforward calculation
\begin{align}
h(\theta,n)= \prod _{m=0}^{\text{N}} \frac{\Gamma \left(\frac{2 m+n+\frac{1}{3}}{2 n}\right)^2 \Gamma \left(\frac{2 m-\frac{i \theta }{\pi }+\frac{5}{3}}{2 n}\right) \Gamma \left(\frac{2 m+2 n+\frac{i \theta }{\pi }+\frac{5}{3}}{2 n}\right)}{\Gamma \left(\frac{2 m+n+\frac{5}{3}}{2
   n}\right)^2 \Gamma \left(\frac{2 m-\frac{i \theta }{\pi }+\frac{1}{3}}{2 n}\right) \Gamma \left(\frac{2 m+2 n+\frac{i \theta }{\pi }+\frac{1}{3}}{2 n}\right)}\times\\
 \exp \l \int_0^\infty \frac{dt}{t\sinh (nt)}e^{-2t(N+1)}\frac{2 \sinh \l \frac{2}{3}t\r }{\sinh t} \sin^2 \l \frac{it}{2}\l n +\frac{i\theta}{\pi}  \r \r \r.
\label{Fmin_mixed} 
\end{align}
For real rapidity difference $\theta$ the integral representation \eqref{Fmin} converges, as we commented before, so the mixed-product form is not really needed. However,  for numerical work, Eq. \eqref{Fmin_mixed} is really useful since the integral factor in \eqref{Fmin_mixed} becomes exponentially suppressed for large $N$.

\begin{figure}[t]
\centering
  \includegraphics[width=0.7\linewidth]{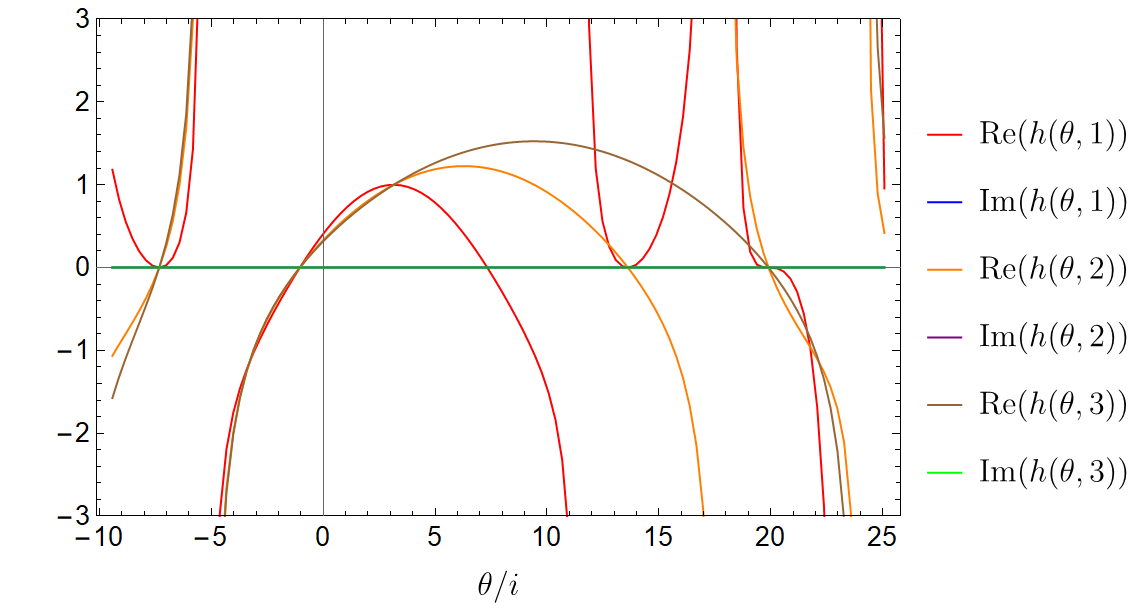}
  \caption{Plot of the minimal form factor in the $n$-th replicated theory for imaginary values of the rapidity. For different values of $n$ ($n=1,2,3$) the real and the imaginary part, which is vanishing, are shown.}
\end{figure}

\section{Computation of $f_{A\bar{A}}(0,\infty,0)$}\label{Analf}tau

We present here an analytic computation of the value of the function $f_{A\bar{A}}(\theta,n,\tau)$ at $\theta=0$, $n=\infty$ and $\tau=0$ in the 3-state Potts model. As seen in Fig.~\ref{f_0_n}  the result is independent of $\tau$, so we simply choose the simplest case $\tau=0$ and follow closely the calculation presented in \cite{Ola} for the sinh-Gordon model. We start from the definition
\beqa
f_{A\bar{A}}(0,n,0) &=& f_{\bar{A} A}(0,n,0)=\frac{1}{\la \mathcal{T}_n\ra^2}  \sum^{n-1}_{j=0} |F_{(A,1)(\bar{A},1)}(2i\pi j,n)|^2 \label{eq_finf}
\eeqa
where the form factor $F_{(A,1)(\bar{A},1)}(\theta)$ is given by
\be
F_{(A,1)(\bar{A},1)}(\theta,n) = \la \mathcal{T}_n\ra\frac{ \sin \frac{\pi}{n} }{2n \sinh \frac{\theta-i\pi}{2n} \sinh\frac{\theta+i\pi}{2n}}\frac{h(\theta,n)}{h(i\pi,n)}.
\ee
Note that this is identical to (\ref{ff}) except for the minimal form factors, which only coincide at $q=3$.

In the sum \eqref{eq_finf}, the $j$-th and $(n-j)$-th terms are identical by periodicity. This means that, in the large $n$ limit, one can replace the sum by
\begin{eqnarray}
f_{A\bar{A}}(0,n,0) &\simeq& \frac{1}{|\la \mathcal{T}_n\ra|^2}\l2 \sum^{\lfloor \frac{n}{2} \rfloor-1}_{j=0} |F_{(A,1)(\bar{A},1)}(2i\pi j,n)|^2 - |F_{(A,1)(\bar{A},1)}(0,n)|^2\r \nonumber \\
&=&\frac{1}{|\la \mathcal{T}_\infty\ra|^2}\l 2 \sum^{\infty}_{j=0} \lim_{n\rightarrow \infty} \left( |F_{(A,1)(\bar{A},1)}(2i\pi j, n)|^2 - |F_{(A,1)(\bar{A},1)}(0, n)|^2 \right)\r.
\end{eqnarray}
The limit inside the sum above can be easily computed. Let us consider the different factors involved in $F_{(A,1)(\bar{A},1)}(\theta,n)$. We have
\be
\frac{ \sin \frac{\pi}{n} }{2n \sinh \frac{\theta-i\pi}{2n} \sinh\frac{\theta+i\pi}{2n}} \rightarrow -\frac{2\pi}{\l 2i\pi j-i\pi\r\l 2i\pi j+i\pi\r},
\ee
and
\be
\frac{h(2i\pi j,n)}{h(i\pi,n)} \rightarrow \frac{\exp\l  -\frac{1}{2}\int_0^\infty \frac{dt}{t} \frac{2\sinh\l \frac{2t}{3} \r }{\sinh t}e^{-2tj}\r}{\exp\l  -\frac{1}{2}\int_0^\infty \frac{dt}{t} \frac{2\sinh\l \frac{2t}{3} \r }{\sinh t}e^{-t}\r} = \frac{\Gamma\l \frac{5}{6}+j\r \Gamma \l 2/3\r}{\Gamma\l \frac{1}{6}+j\r \Gamma(4/3)}.
\ee
Putting all the pieces together, we get
\be
f_{A\bar{A}}(0,\infty,0) =\frac{8}{\pi^2}\sum^{\infty}_{j=0} \frac{1}{(2j-1)^2(2j+1)^2}\frac{\Gamma^2\l \frac{5}{6}+j\r \Gamma^2\l \frac{3}{2}\r}{\Gamma^2\l \frac{1}{6}+j\r\Gamma^2\l \frac{4}{3}\r}-\frac{4}{\pi^2}\frac{\Gamma^2\l \frac{5}{6}\r \Gamma^2\l \frac{3}{2}\r}{\Gamma^2\l \frac{1}{6}\r\Gamma^2\l \frac{4}{3}\r}.
\ee
The result is given in terms generalized hypergeometric functions defined as
\begin{equation}
   _{p}F_{q}\left[ \begin{array}{c}
     a_1, a_2, \ldots, a_p ; z \\
     b_1, b_2, \ldots, b_q \\
   \end{array}\right]=\sum_{k=0}^{\infty} \frac{(a_1)_k (a_2)_k \ldots (a_p)_k}{(b_1)_k (b_2)_k \ldots
   (b_q)_k}\frac{z^k}{k!},
\end{equation}
where $(a)_k \equiv \frac{\Gamma(a+k)}{\Gamma(a)}$ is the Pochhammer symbol. The sum of both particle orderings $f_{A\bar{A}}(0,\infty,0) + f_{\bar{A}A}(0,\infty,0)=2f_{A\bar{A}}(0,\infty,0)$ then gives
\begin{eqnarray}
2f_{A\bar{A}}(0,\infty,0) = \frac{16}{\pi^2}\frac{\Gamma^2\l 5/6\r \Gamma^2\l 2/3\r}{\Gamma^2\l 4/3\r \Gamma^2\l 1/6\r}\l    _{5}F_{4}\left[ \begin{array}{c}
     -1/2, -1/2, 5/6, 5/6, 1 ; 1 \\
    1/2, 1/2, 3/2, 3/2 \\
   \end{array}\right] -1/2\r  \simeq  0.575153
\end{eqnarray}

\end{appendix}


\begin{thebibliography}{99}


\bibitem{vNE1} L.~Amico, R.~Fazio, A.~Osterloh, and V.~Vedral,
\emph{Entanglement in many-body systems}, \href{http://dx.doi.org/10.1103/RevModPhys.80.517}{Rev. Mod. Phys. \textbf{80}, 517 (2008)}.


\bibitem{vNE2} P.~Calabrese, J. Cardy, and B.~Doyon, \emph{Entanglement
entropy in extended quantum systems}, \href{http://dx.doi.org/10.1088/1751-8121/42/50/500301}{J. Phys. A \textbf{42}, 500301 (2009)}.

\bibitem{vNE3} J.~Eisert, M.~Cramer, and M.~B.~Plenio, \emph{Area
laws for the entanglement entropy}, \href{http://dx.doi.org/10.1103/RevModPhys.82.277}{Rev. Mod. Phys. \textbf{82}, 277 (2010)}.

\bibitem{vNE4} N.~Laflorencie, \textit{Quantum entanglement in condensed
matter systems}, \href{http://dx.doi.org/10.1016/j.physrep.2016.06.008}{Phys. Rep. \textbf{643}, 1 (2016)}.

 \bibitem{cl-08}
P. Calabrese and A. Lefevre,
{\it Entanglement spectrum in one-dimensional systems},
 \href{http://dx.doi.org/10.1103/PhysRevA.78.032329}{Phys. Rev. A {\bf 78}, 032329 (2008)}.
 
 
 \bibitem{ace-18}V. Alba, P. Calabrese, and E. Tonni, {\it Entanglement spectrum degeneracy and Cardy formula in 1+1 dimensional conformal field theories},
\href{https://doi.org/10.1088/1751-8121/aa9365}{J. Phys.  A {\bf 51}, 024001 (2018)}.

\bibitem{cc-04} P. Calabrese and J. Cardy, \textit{Entanglement Entropy and Quantum Field Theory}, \href{https://iopscience.iop.org/article/10.1088/1742-5468/2004/06/P06002}{J. Stat. Mech. P06002 (2004).} 

\bibitem{gs-18} M. Goldstein and E. Sela, \textit{Symmetry-Resolved
Entanglement in Many-Body Systems}, \href{http://dx.doi.org/10.1103/PhysRevLett.120.200602}{Phys. Rev. Lett. \textbf{120}, 200602 (2018)}.

\bibitem{lr-14} N. Laflorencie and S. Rachel, \textit{Spin-resolved
entanglement spectroscopy of critical spin chains and Luttinger liquids},
\href{http://dx.doi.org/10.1088/1742-5468/2014/11/P11013}{J. Stat. Mech. P11013 (2014)}.

\bibitem{Greiner} A. Lukin, M. Rispoli, R. Schittko, M. E. Tai, A.
M. Kaufman, S. Choi, V. Khemani, J. Leonard, and M. Greiner, \textit{Probing
entanglement in a many-body localized system}, \href{https://dx.doi.org/10.1126/science.aau0818}{Science \textbf{364}, 6437 (2019)}.

\bibitem{SRENegativity} E. Cornfeld, M. Goldstein, and E. Sela, \textit{Imbalance
Entanglement: Symmetry Decomposition of Negativity}, \href{http://dx.doi.org/10.1103/PhysRevA.98.032302}{Phys. Rev. A \textbf{98}, 032302 (2018)}.

\bibitem{Equipartitioning} J. C. Xavier, F. C. Alcaraz, and G. Sierra,
\textit{Equipartition of the entanglement entropy}, \href{https://journals.aps.org/prb/abstract/10.1103/PhysRevB.98.041106}{Phys. Rev. B \textbf{98}, 041106 (2018)}.

\bibitem{SREQuench} N. Feldman and M. Goldstein, \textit{Dynamics
of Charge-Resolved Entanglement after a Local Quench}, \href{http://dx.doi.org/10.1103/PhysRevB.100.235146}{Phys. Rev. B \textbf{100}, 235146 (2019)}.

\bibitem{bc-20} R. Bonsignori and P. Calabrese, \textit{Boundary
effects on symmetry resolved entanglement}, \href{https://doi.org/10.1088/1751-8121/abcc3a}{J. Phys. A  \textbf{54}, 015005 (2021)}.

\bibitem{crc-20} L. Capizzi, P. Ruggiero, and P. Calabrese, \textit{Symmetry
resolved entanglement entropy of excited states in a CFT}, \href{ttp://dx.doi.org/10.1088/1742-5468/ab96b6}{J. Stat. Mech. (2020) 073101}.

\bibitem{mbc-21} S. Murciano, R. Bonsignori, and P. Calabrese, \textit{Symmetry
decomposition of negativity of massless free fermions}, \href{https://arxiv.org/abs/2102.10054}{arXiv:2102.10054 (2021)}.

\bibitem{Chen-21}H.-H. Chen,  \textit{Symmetry decomposition of relative entropies in conformal field theory},
\href{https://arxiv.org/abs/2104.03102}{arXiv:2104.03102.}

\bibitem{cc-21}L. Capizzi and P. Calabrese,  \textit{Symmetry resolved relative entropies and distances in conformal field theory,}
\href{https://arxiv.org/abs/2105.08596}{arXiv:2105.08596.}

\bibitem{cdmWZW-21}P. Calabrese, J. Dubail, and S. Murciano,  \textit{Symmetry-resolved entanglement entropy in Wess-Zumino-Witten models},
\href{https://arxiv.org/pdf/2106.15946.pdf}{arXiv:2106.15946.}

\bibitem{mdc-20b} S. Murciano, G. Di Giulio, and P. Calabrese, \textit{Entanglement
and symmetry resolution in two dimensional free quantum field theories},
\href{https://doi.org/10.1007/JHEP08(2020)073}{JHEP \textbf{08} 073 (2020)}.

\bibitem{U(1)FreeFF}D. X. Horváth, L. Capizzi, and P. Calabrese,
\textit{U(1) symmetry resolved entanglement in free 1+1 dimensional
field theories via form factor bootstrap}, \href{https://doi.org/10.1007/JHEP05(2021)197}{JHEP {\bf 05}, 197 (2021).}

\bibitem{Z2IsingShg}D. X. Horváth and P. Calabrese, \textit{Symmetry
resolved entanglement in integrable field theories via form factor
bootstrap}, \href{https://doi.org/10.1007/JHEP11(2020)131}{JHEP \textbf{11}, 131 (2020).}

\bibitem{SGSRE} D. X. Horváth,  P. Calabrese, and O. A. Castro-Alvaredo, \textit{Branch Point Twist Field Form Factors in the sine-Gordon Model II: 
Composite Twist Fields and Symmetry Resolved Entanglement}, ArXiv preprint  \href{https://arxiv.org/abs/2105.13982}{2105.13982}.

\bibitem{znm}S. Zhao, C. Northe, and R. Meyer, \textit{Symmetry-Resolved
Entanglement in AdS3/CFT2 coupled to U(1) Chern-Simons Theory, }\href{https://arxiv.org/abs/2012.11274}{arXiv:2012.11274}.

\bibitem{znm2}K. Weisenberger, S. Zhao, C. Northe, and R. Meyer, \textit{Symmetry-resolved entanglement for excited states and two entangling intervals in AdS$_3$/CFT$_2$, }\href{https://arxiv.org/abs/2108.09210}{arXiv:2108.09210}.

\bibitem{brc-19} R.~Bonsignori, P. Ruggiero, and P. Calabrese, \textit{Symmetry
resolved entanglement in free fermionic systems}, \href{https://doi.org/10.1088/1751-8121/ab4b77}{J. Phys. A \textbf{52}, 475302 (2019)}.

\bibitem{fg-20} S. Fraenkel and M. Goldstein, \textit{Symmetry resolved
entanglement: Exact results in 1d and beyond}, \href{http://dx.doi.org/10.1088/1742-5468/ab7753}{J. Stat. Mech. 033106 (2020)}.

\bibitem{FreeF1} H. Barghathi, C. M. Herdman, and A. Del Maestro,
\textit{Rényi Generalization of the Accessible Entanglement Entropy},
\href{https://doi.org/10.1103/PhysRevLett.121.150501}{Phys. Rev. Lett. \textbf{121}, 150501 (2018)}.

\bibitem{FreeF2} H. Barghathi, E. Casiano-Diaz, and A. Del Maestro,
\textit{Operationally accessible entanglement of one dimensional spinless
fermions}, \href{https://doi.org/10.1103/PhysRevA.100.022324}{Phys. Rev. A \textbf{100}, 022324 (2019)}.

\bibitem{mdc-20} S. Murciano, G. Di Giulio, and P. Calabrese, \textit{Symmetry
resolved entanglement in gapped integrable systems: a corner transfer
matrix approach}, \href{https://dx.doi.org/10.21468/SciPostPhys.8.3.046}{SciPost Phys. \textbf{8}, 046 (2020)}.

\bibitem{ccdm-20} P. Calabrese, M. Collura, G. Di Giulio, and S.
Murciano, \textit{Full counting statistics in the gapped XXZ spin
chain}, \href{https://doi.org/10.1209/0295-5075/129/60007}{EPL \textbf{129}, 60007 (2020)}.

\bibitem{pbc-20} G. Parez, R. Bonsignori, and P. Calabrese, \textit{Quasiparticle
dynamics of symmetry resolved entanglement after a quench: the examples
of conformal field theories and free fermions}, \href{https://doi.org/10.1103/PhysRevB.103.L041104}{Phys. Rev. B \textbf{103}, L041104  (2021)}.

\bibitem{HigherDimFermions} M. T. Tan and S. Ryu, \textit{Particle
Number Fluctuations, Rényi and Symmetry-resolved Entanglement Entropy
in Two-dimensional Fermi Gas from Multi-dimensional bosonisation},
\href{https://dx.doi.org/10.1103/PhysRevB.101.235169}{Phys. Rev. B \textbf{101}, 235169 (2020)}.

\bibitem{mrc-20} S. Murciano, P. Ruggiero, and P. Calabrese, \textit{Symmetry
resolved entanglement in two-dimensional systems via dimensional reduction},
\href{https://dx.doi.org/10.1088/1742-5468/aba1e5}{J. Stat. Mech. 083102 (2020)}.

\bibitem{ncv-21}
A. Neven, J. Carrasco, V. Vitale, C. Kokail, A. Elben, M. Dalmonte, P. Calabrese, P. Zoller, B. Vermersch, R. Kueng, and B. Kraus,
{\it Symmetry-resolved entanglement detection using partial transpose moments},  \href{https://arxiv.org/abs/2103.07443}{arXiv:2103.07443}.

\bibitem{fg-21}
S. Fraenkel and M. Goldstein, {\it Entanglement Measures in a Nonequilibrium Steady State: Exact Results in One Dimension},
 \href{https://arxiv.org/abs/2105.00740}{arXiv:2105.00740}.
 
 \bibitem{pbc-21}
G. Parez, R. Bonsignori, and P. Calabrese, {\it Exact quench dynamics of symmetry resolved entanglement in a free fermion chain},
 \href{https://arxiv.org/pdf/2106.13115.pdf}{arXiv:2106.13115}.

\bibitem{trac-20} X. Turkeshi, P. Ruggiero, V. Alba, and P. Calabrese,
\textit{Entanglement equipartition in critical random spin chains},
\href{https://doi.org/10.1103/PhysRevB.102.014455}{Phys. Rev. B \textbf{102}, 014455 (2020)}.

\bibitem{MBL} M. Kiefer-Emmanouilidis, R. Unanyan, J. Sirker, and
M. Fleischhauer, \textit{Evidence for unbounded growth of the number
entropy in many-body localized phases}, \href{https://dx.doi.org/10.1103/PhysRevLett.124.243601}{Phys. Rev. Lett. \textbf{124}, 243601 (2020)}.

\bibitem{MBL2} M. Kiefer-Emmanouilidis, R. Unanyan, M. Fleischhauer,
and J. Sirker \textit{Unlimited growth of particle fluctuations in
many-body localized phases}, \href{https://doi.org/10.1016/j.aop.2021.168481}{Ann. Phys. 168481 (2021)}.

\bibitem{Topology} K. Monkman and J. Sirker, \textit{Operational
Entanglement of Symmetry-Protected Topological Edge States}, \href{https://doi.org/10.1103/PhysRevResearch.2.043191}{Phys. Rev. Res. \textbf{2}, 043191 (2020)}.

\bibitem{Anyons} E. Cornfeld, L. A. Landau, K. Shtengel, and E. Sela,
\textit{Entanglement spectroscopy of non-Abelian anyons: Reading off
quantum dimensions of individual anyons}, \href{http://dx.doi.org/10.1103/PhysRevB.99.115429}{Phys. Rev. B \textbf{99}, 115429 (2019)}.

\bibitem{as-20} D. Azses and E. Sela, 
\textit{Symmetry-resolved entanglement in symmetry-protected topological phases},
\href{https://doi.org/10.1103/PhysRevB.102.235157}{Phys. Rev. B \textbf{102}, 235157 (2020).}

\bibitem{vecd-20} V. Vitale, A. Elben, R. Kueng, A. Neven, J. Carrasco,
B. Kraus, P. Zoller, P. Calabrese, B. Vermersch, and M. Dalmonte,
\textit{Symmetry-resolved dynamical purification in synthetic quantum
matter}, \href{https://arxiv.org/abs/2101.07814}{ArXiv:2101.07814}.



\bibitem{CZam} L. Chim and A. B. Zamolodchikov, \textit{Integrable field theory of $q$-state Potts model with $0 < q < 4$}, \href{https://doi.org/10.1142/S0217751X9200243X} {Int. J. Mod. Phys. A {\bf 7}, 5317 (1992)}.

\bibitem{DoPoTa1} P. Dorey, A. Pocklington, and R. Tateo, \textit{Integrable aspects of the scaling q state Potts models. 1. Bound states and bootstrap closure}, \href{https://doi.org/10.1016/S0550-3213(03)00181-0}{Nucl. Phys. B {\bf 661}, 425 (2003)}.

\bibitem{DoPoTa2}  P. Dorey, A. Pocklington, and R. Tateo, \textit{Integrable aspects of the scaling q state Potts models. 2. Finite size effects}, \href{https://doi.org/10.1016/S0550-3213(03)00182-2}{Nucl. Phys. B {\bf B661}, 464 (2003)}.

\bibitem{SaJiMi} M. Sato, M. Jimbo and T. Miwa, \textit{Studies on Holonomic Quantum Fields I}, \href{10.3792/pjaa.53.6}{Proc. Japan. Acad. Sci. {\bf A53} (1977)}.

\bibitem{ZaPotts}  A. Zamolodchikov, \textit{Integrals of motion in scaling 3-state Potts model field theory}, \href{https://doi.org/10.1142/S0217751X88000333} {Int. J. Mod. Phys. A {\bf 3}, 743 (1988)}.

\bibitem{ZZ} A. B. Zamolodchikov and A. B. Zamolodchikov, \textit{Factorized
$S$-matrices in two dimensions as the exact solutions of certain
relativistic quantum field theory models,} \href{https://doi.org/10.1016/0003-4916(79)90391-9}{Ann. Phys. \textbf{120}, 253 (1979).}

\bibitem{FA} L. D. Faddeev, \textit{Quantum completely integrable models in field theory}, \href{https://doi.org/10.1142/9789812815453_0007}{Cont. Math. Phys., {\bf 1C} (1980) 107}. 

\bibitem{SmirnovBook}F. Smirnov, \textit{ Form factors in completely
integrable models of quantum field theory}, \href{https://doi.org/10.1142/1115}{Adv. Series in Math. Phys. 14, World Scientific, Singapore (1992).}

\bibitem{KarowskiU1}M. Karowski and P. Weisz, \textit{Exact Form-Factors
in (1+1)-Dimensional Field Theoretic Models With Soliton Behavior,}
\href{https://doi.org/10.1016/0550-3213(78)90362-0}{Nucl. Phys. B \textbf{139}, 455 (1978)}.

\bibitem{k-87} V. Knizhnik, \textit{Analytic fields on riemann surfaces.
II}, \href{http://dx.doi.org/10.1007/BF01225373}{Comm. Math. Phys. \textbf{112},  567 (1987)}.

\bibitem{dixon} L. J. Dixon, D. Friedan, E. J. Martinec, and S. H.
Shenker, \textit{The Conformal Field Theory Of Orbifolds}, \href{https://doi.org/10.1016/0550-3213(87)90676-6}{ Nucl. Phys. B \textbf{282}, 13 (1987)}.


\bibitem{Baxter}
R.J. Baxter, \textit{Exactly solved models of statistical mechanics}, Academic Press, London, 1982.

\bibitem{pagialle}
P. Di Francesco, P. Mathieu, and D. Senechal, {\it Conformal Field Theory} (Springer-Verlag, New York, 1997).

\bibitem{Muss}
 G. Mussardo, {\it Statistical field theory: an introduction to exactly solved models in statistical
physics}, 2nd edition, Oxford University Press (2020).


\bibitem{Aferr_Delf}
G. Delfino, \textit{Sine-Gordon description of the scaling three state Potts antiferromagnet on the square lattice}. \href{https://doi.org/10.1088/0305-4470/34/21/102}{J. Phys.  A \textbf{34}, L311  (2001)}.

\bibitem{Potts_Delf}
G. Delfino J.L.Cardy. \textit{Universal amplitude ratios in the two-dimensional q-state Potts model and percolation from quantum field theory}. \href{https://doi.org/10.1016/S0550-3213(98)00144-8}{Nucl. Phys. B {\bf 519}, 551 (1998) 551}.

\bibitem{delta_theorem} G. Delfino, P. Simonetti, and J. L. Cardy,
\textit{Asymptotic factorisation of form factors in two-dimensional
quantum field theory}, \href{https://doi.org/10.1016/0370-2693(96)01035-0}{Phys. Lett. B \textbf{387}, 327 (1996)}.

\bibitem{Dis_Delfino}
G. Delfino, \textit{Particles, conformal invariance and criticality in pure and disordered systems}. \href{https://doi.org/10.1140/epjb/s10051-021-00076-0}{Eur. Phys. J. B \textbf{94}, 65 (2021)}.

\bibitem{Part_Delfino}
G. Delfino, \textit{Fields, particles and universality in two dimensions}, \href{https://doi.org/10.1016/j.aop.2015.05.020}{Ann. Phys. {\bf 360},  477 (2015)}.


\bibitem{Zn_integ}
R. K\"{o}berle, J. A. Swieca. \textit{Factorizable Z(N) models}, \href{https://doi.org/10.1016/0370-2693(79)90822-0}{Phys. Lett. B {\bf 86}, 209 (1979)}.

\bibitem{Zn_Zam}
V. A. Fateev, A. B. Zamolodchikov, \textit{Self-dual solutions of the star-triangle relations in $\mathbb{Z}_N$-models}. \href{https://doi.org/10.1016/0375-9601(82)90736-8}{Phys. Lett. A {\bf 92}, 37 (1982).}

\bibitem{Zn_Zam2}
A.B. Zamolodchikov and V.A. Fateev. \textit{Nonlocal (parafermion) currents in two-dimensional conformal quantum field theory and self-dual critical points in $\mathbb{Z}_N$--symmetric statistical
systems}, Zh. Eksper. Teoret. Fiz., {\bf{89}}, 380 (1985).

\bibitem{Zn_Parafermions}
V. A. Fateev, V. V. Postnikov, and Y. P. Pugai, \textit{On scaling fields in $\mathbb{Z}_N$ Ising models}, \href{https://doi.org/10.1134/S0021364006040096}{JETP Lett. \textbf{83}, 172, (2006)}.

\bibitem{fz-91}
V.A.Fateev, Al.B. Zamolodchikov, \textit{Integrable perturbations of $\mathbb{Z}_N$ parafermion models and the O(3) sigma model}, \href{https://doi.org/10.1016/0370-2693(91)91283-2}{Phys. Lett. B {\bf 271}, 91 (1991) }.

\bibitem{Ola}
J.L. Cardy, O.A. Castro-Alvaredo and B. Doyon, \textit{Form factors of branch-point twist fields in quantum integrable models and entanglement entropy}. \href{https://doi.org/10.1007/s10955-007-9422-x}{J. Stat. Phys. \textbf{130}, 129 (2008)}.

\bibitem{cd-08} O. A. Castro-Alvaredo and B. Doyon, \textit{Bi-partite
entanglement entropy in integrable models with backscattering}, \href{http://dx.doi.org/10.1088/1751-8113/41/27/275203}{J. Phys. A \textbf{41},  275203 (2008)}.

\bibitem{Ola1}
O. A. Castro-Alvaredo, \textit{Massive Corrections to Entanglement in Minimal E8 Toda Field Theory}. \href{https://doi.org/10.21468/SciPostPhys.2.1.008}{SciPost Phys. {\bf 2}, 008 (2017)}

\bibitem{Ola2}
D. Bianchini, O. A. Castro-Alvaredo, \textit{Branch Point Twist Field Correlators in the Massive Free Boson Theory}. \href{https://doi.org/10.1016/j.nuclphysb.2016.10.016}{Nucl. Phys. B \bf{913}, 879 (2016)}.

\bibitem{next}B. Doyon, \textit{Bi-partite entanglement entropy in massive
two-dimensional quantum field theory}, \href{https://doi.org/10.1103/PhysRevLett.102.031602}{Phys. Rev. Lett. \textbf{102}, 031602 (2009)}.

\bibitem{cd-09} O. A. Castro-Alvaredo and B. Doyon, \textit{Bi-partite
entanglement entropy in massive 1+1-dimensional quantum field theories},
\href{http://dx.doi.org/10.1088/1751-8113/42/50/504006}{J. Phys. A \textbf{42}, 504006 (2009)}.

\bibitem{cd-09b} O. A. Castro-Alvaredo and B. Doyon, \textit{Bi-partite
entanglement entropy in massive QFT with a boundary: the Ising model},
\href{http://dx.doi.org/10.1007/s10955-008-9664-2}{J. Stat. Phys.  \textbf{134}, 105 (2009)}.

\bibitem{cl-11} O. A. Castro-Alvaredo and E. Levi, \textit{Higher
particle form factors of branch point twist fields in integrable quantum
field theories}, \href{https://doi.org/10.1088/1751-8113/44/25/255401}{J. Phys. A \textbf{44} (2011) 255401}.

\bibitem{cdl-12} O. A. Castro-Alvaredo, B. Doyon, and E. Levi, \textit{Arguments
towards a c-theorem from branch-point twist fields}, \href{https://doi.org/10.1088/1751-8113/44/49/492003}{J. Phys.  A \textbf{44} (2011) 492003}.

\bibitem{leviFFandVEV} E. Levi,\textit{ Composite branch-point twist
fields in the Ising model and their expectation values, }\href{https://doi.org/10.1088/1751-8113/45/27/275401}{J. Phys. A  \textbf{45} (2012) 275401}.

\bibitem{lcd-13} E. Levi, O. A. Castro-Alvaredo, and B. Doyon, \textit{Universal
corrections to the entanglement entropy in gapped quantum spin chains:
a numerical study}, \href{http://dx.doi.org/10.1103/PhysRevB.88.094439}{Phys. Rev. B \textbf{88}, 094439 (2013)}.

\bibitem{bcd-15} D. Bianchini, O. Castro-Alvaredo, B. Doyon, E. Levi,
and F. Ravanini, \textit{Entanglement Entropy of Non Unitary Conformal
Field Theory}, \href{https://doi.org/10.1088/1751-8113/48/4/04FT01}{ J. Phys. A \textbf{48}, 04FT01 (2014)}.

\bibitem{bcd-15b} D. Bianchini, O. Castro-Alvaredo, and B. Doyon,
\textit{Entanglement Entropy of Non-Unitary Integrable Quantum Field
Theory}, \href{https://doi.org/10.1016/j.nuclphysb.2015.05.013}{Nucl. Phys.  B \textbf{896} (2015) 835}.

\bibitem{bcd-16}O. Blondeau-Fournier, O. A. Castro-Alvaredo, and
B. Doyon, \textit{Universal scaling of the logarithmic negativity
in massive quantum field theory}, \href{https://doi.org/10.1088/1751-8113/49/12/125401}{J. Phys. A \textbf{49}, 125401 (2016)}.


\bibitem{clsv-19} O. A. Castro-Alvaredo, M. Lencs\'es, I. M. Sz\'ecs\'enyi,
and J. Viti, \textit{Entanglement Dynamics after a Quench in Ising
Field Theory: A Branch Point Twist Field Approach}, \href{http://dx.doi.org/10.1007/JHEP12(2019)079}{JHEP \textbf{12} (2019) 79}.

\bibitem{Ola3}
O. A. Castro-Alvaredo, M. Lencs\'es, I. M. Sz\'ecs\'enyi, J. Viti, \textit{Entanglement Oscillations near a Quantum Critical Point}. \href{https://doi.org/10.1103/PhysRevLett.124.230601}{Phys. Rev. Lett. {\bf 124}, 230601 (2020)}.

\bibitem{Ola-c}
O. A. Castro-Alvaredo, C. De Fazio, B. Doyon, I. M. Sz\'ecs\'enyi, \textit{Entanglement Content of Quasi-Particle Excitations}. \href{https://doi.org/10.1103/PhysRevLett.121.170602}{Phys. Rev. Lett. {\bf 121}, 170602 (2018)}.

\bibitem{Ola-c1}
O. A. Castro-Alvaredo, C. De Fazio, B. Doyon, I. M. Sz\'ecs\'enyi, \textit{Entanglement Content of Quantum Particle Excitations I. Free Field Theory}. \href{https://doi.org/10.1007/JHEP10(2018)039}{JHEP {\bf 10} (2018) 039}.

\bibitem{Ola-c2}
O. A. Castro-Alvaredo, C. De Fazio, B. Doyon, I. M. Sz\'ecs\'enyi, \textit{Entanglement Content of Quantum Particle Excitations II. Disconnected Regions and Logarithmic Negativity}. \href{https://doi.org/10.1007/JHEP11(2019)058}{JHEP {\bf 11} (2019) 58}.

\bibitem{Ola-c3}
O. A. Castro-Alvaredo, C. De Fazio, B. Doyon, I. M. Sz\'ecs\'enyi, \textit{Entanglement Content of Quantum Particle Excitations III. Graph Partition Functions}. \href{https://doi.org/10.1063/1.5098892}{J. Math. Phys. {\bf 60}, 082301 (2019)}.


\bibitem{Zn}
B. Estienne, Y. Ikhlef, A. Morin-Duchesne, \textit{Finite-size corrections in critical symmetry-resolved entanglement}. \href{https://doi.org/10.21468/SciPostPhys.10.3.054}{SciPost Phys. {\bf 10}, 054 (2021)}








\bibitem{bym-13} A. Belin, L.-Y. Hung, A. Maloney, S. Matsuura, R.
C. Myers, and T. Sierens, \textit{Holographic charged R\'enyi entropies},
\href{https://doi.org/10.1007/JHEP12(2013)059}{ JHEP \textbf{12} (2013) 059}.

\bibitem{cms-13} P. Caputa, G. Mandal, and R. Sinha, \textit{Dynamical
entanglement entropy with angular momentum and U(1) charge}, \href{https://dx.doi.org/10.1007/JHEP11(2013)052}{JHEP \textbf{11} (2013) 052}.

\bibitem{cnn-16} P. Caputa, M. Nozaki, and T. Numasawa, \textit{Charged
Entanglement Entropy of Local Operators}, \href{https://dx.doi.org/10.1103/PhysRevD.93.105032}{Phys. Rev. D \textbf{93}, 105032 (2016)}.

\bibitem{d-16} J. S. Dowker, \textit{Conformal weights of charged
R\'enyi entropy twist operators for free scalar fields in arbitrary
dimensions}, \href{https://doi.org/10.1088/1751-8113/49/14/145401}{J. Phys. A \textbf{ 49}, 145401 (2016)}.

\bibitem{d-16b} J. S. Dowker, \textit{Charged R\'enyi entropies for
free scalar fields}, \href{https://doi.org/10.1088/1751-8121/aa6178}{ J. Phys. A \textbf{ 50}, 165401 (2017)}.

\bibitem{ssr-17} H. Shapourian, K. Shiozaki, and S. Ryu, \textit{Partial
time-reversal transformation and entanglement negativity in fermionic
systems}, \href{https://doi.org/10.1103/PhysRevB.95.165101}{Phys. Rev. B \textbf{ 95}, 165101 (2017)}.

\bibitem{srrc-19} H. Shapourian, P. Ruggiero, S. Ryu, and P. Calabrese,
\textit{Twisted and untwisted negativity spectrum of free fermions},
\href{https://doi.org/10.21468/SciPostPhys.7.3.037}{SciPost Phys. \textbf{7}, 037 (2019)}.


\end{thebibliography}
\end{document}